\documentclass[aps,prb,amsmath,twocolumn,amssymb,floatfixng,showpacs,
superscriptaddress,footinbib,multicol]{revtex4-1}
\usepackage{graphicx}
\usepackage{epstopdf}
\usepackage{dcolumn}
\usepackage{bm}
\usepackage{color}
\usepackage{amsmath}
\usepackage{braket}
\usepackage{amsfonts}
\usepackage{soul}
\usepackage{url}
\usepackage{array,tabularx}
\usepackage{graphics}
\usepackage{times}
\usepackage{subfigure}
\usepackage[export]{adjustbox}
\usepackage[colorlinks=true,linktoc=page,linkcolor=red,citecolor=blue]{hyperref}
\newcommand\x{4.8}
\newcommand\y{2.5}

\begin{document}

\title{Controlled fast wavepackets in non-Hermitian lattices}
\author{Yehonatan Benisty}
\affiliation{School of Mechanical Engineering, Tel Aviv University, Tel Aviv 69978, Israel}
\author{Sayan Jana}
\affiliation{School of Mechanical Engineering, Tel Aviv University, Tel Aviv 69978, Israel}
\author{Lea Sirota}\email{leabeilkin@tauex.tau.ac.il}
\affiliation{School of Mechanical Engineering, Tel Aviv University, Tel Aviv 69978, Israel}

\begin{abstract}

We report the propagation of fast wavepackets in classical non-Hermitian lattices, where the group velocity is controlled by the non-Hermiticity parameters, and can be made higher than in the Hermitian counterpart. 
Specifically, we obtain a square root dependence of the group velocity on the gain/loss parameter, similarly to the dependence of quantum wavepackets in stretched graphene-like lattices subjected to gain and loss. 
We derive a targeted mapping from the quantum to the classical Hamiltonian to realize this phenomenon in a dynamically stable form. As a result, fast wavepackets of any frequency supported by the lattice are propagating in time domain with a non-growing amplitude.
We demonstrate the system experimentally in a topoelectrical metamaterial, where the non-Hermiticity is generated by embedded operational amplifiers in a feedback setup. 
Our design paves the way to realize increased group velocities, and other wave phenomena inspired by quantum systems in a form that preserves the original system properties, while supporting an inherently stable dynamics. 

\end{abstract}

\maketitle

\begin{figure*}[htpb]
    \centering 
    \begin{tabular}[t]{c|c}
    \begin{tabular}{c}
    \textbf{(a)} \\ 
     \includegraphics[width=5.8 cm, valign=t]{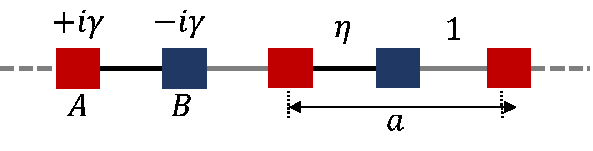} \\
     \textbf{(c)} $\Box$ \\ 
    \begin{tabular}{ccc}
       \includegraphics[height=\y cm, valign=c]{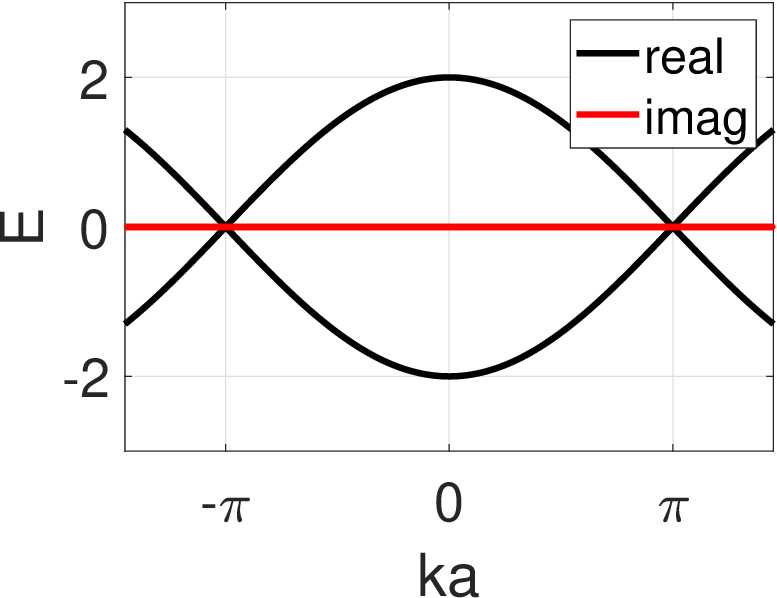}    &  \includegraphics[height=\y cm, valign=c]{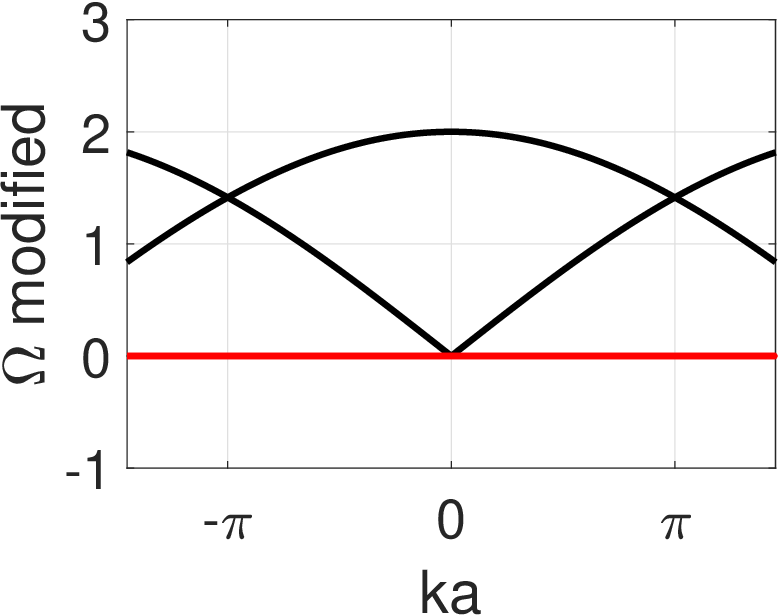} &  \includegraphics[height=\y cm, valign=c]{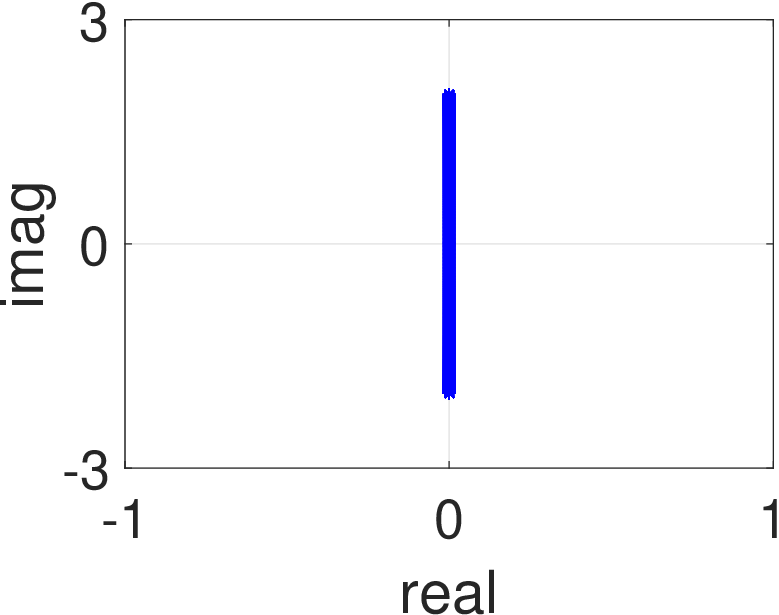} 
    \end{tabular}
    \end{tabular} & \begin{tabular}{cc}
         \textbf{(b)} &
    \includegraphics[width=5.9 cm, valign=t]{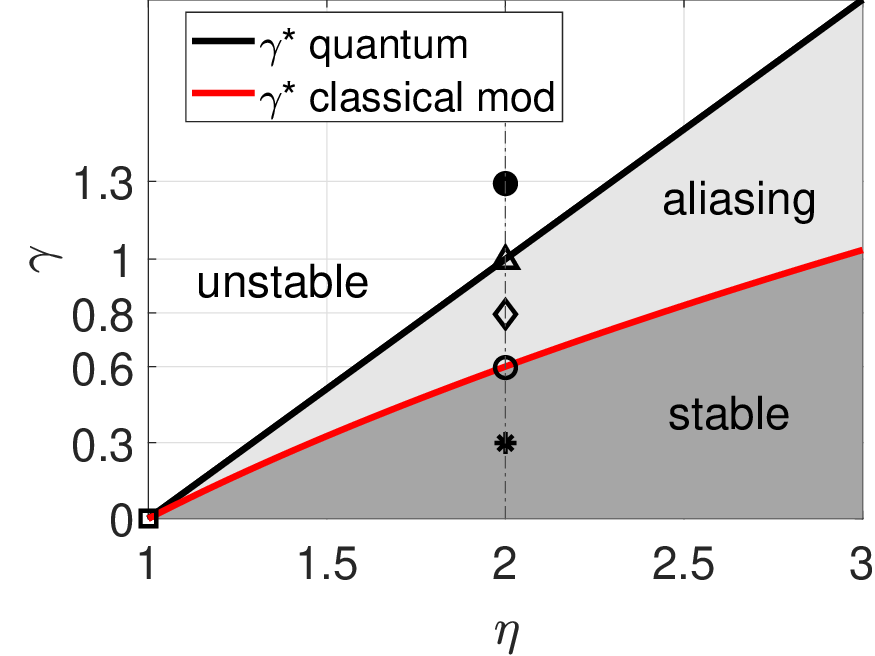} 
    \end{tabular}
    \end{tabular} 
\\
 \begin{tabular}{c c c c c}
   \textbf{(d)} $\bullet$  &  \textbf{(e)} $\triangle$  &   \textbf{(f)} $\Diamond$  & \textbf{(g)} $\circ$  & \textbf{(h)} $*$   \\
   \includegraphics[height=\y cm, valign=c]{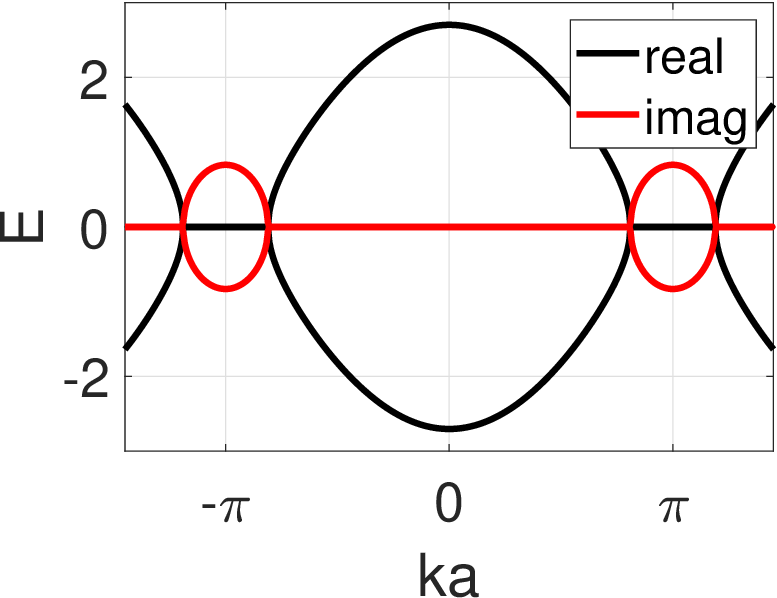}    & 
    \includegraphics[height=\y cm, valign=c]{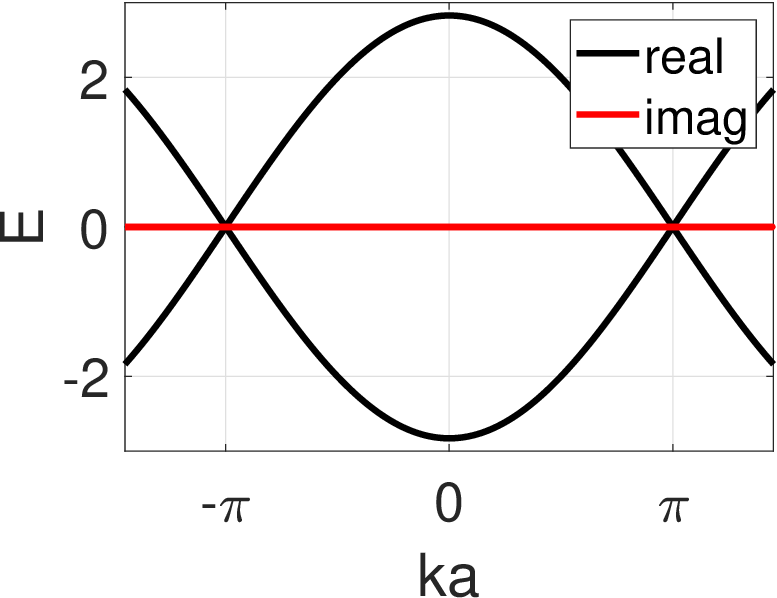}    &\includegraphics[height=\y cm, valign=c]{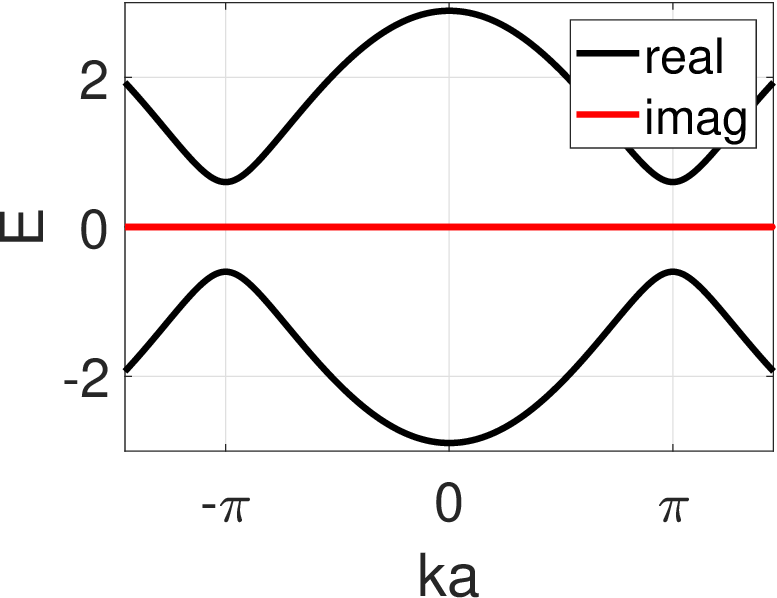} &  \includegraphics[height=\y cm, valign=c]{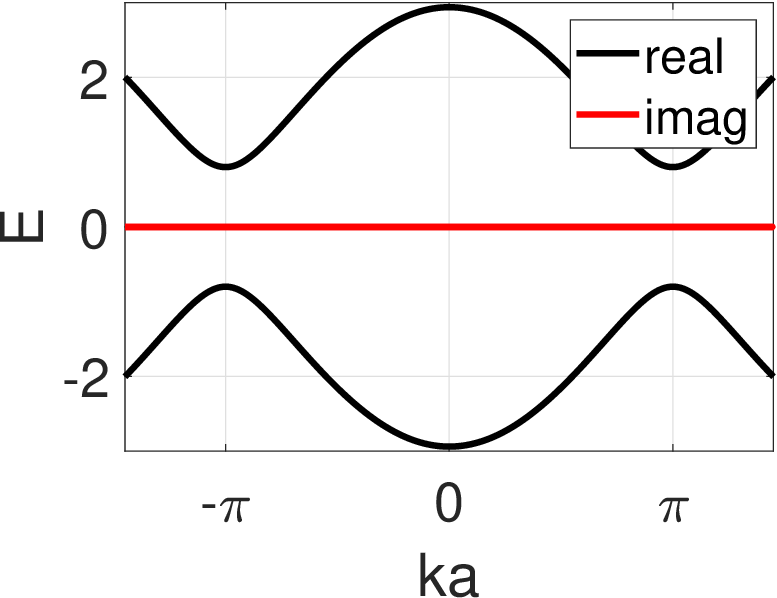} & \includegraphics[height=\y cm, valign=c]{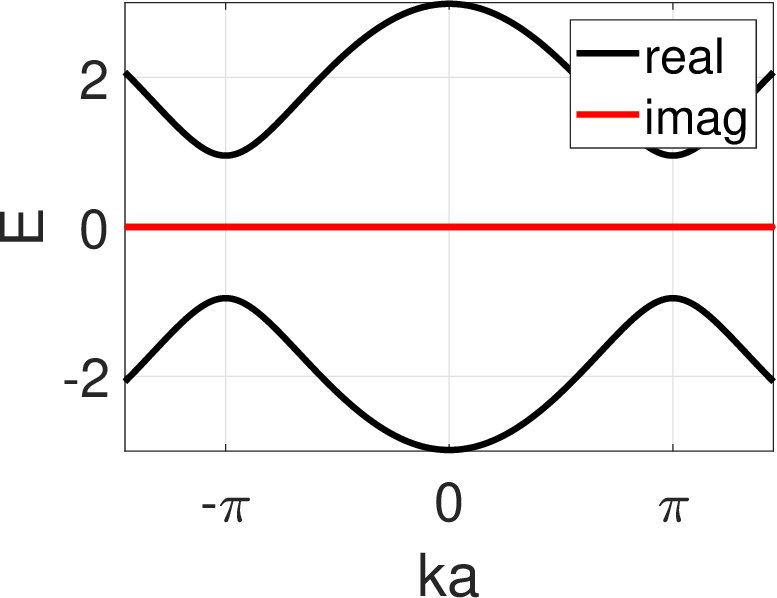}
 \\
  \includegraphics[height=\y cm, valign=c]{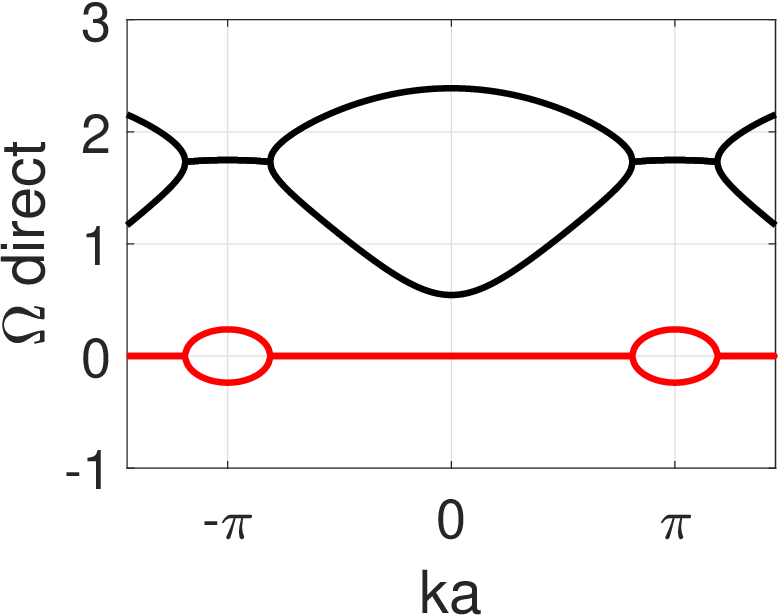}    & 
    \includegraphics[height=\y cm, valign=c]{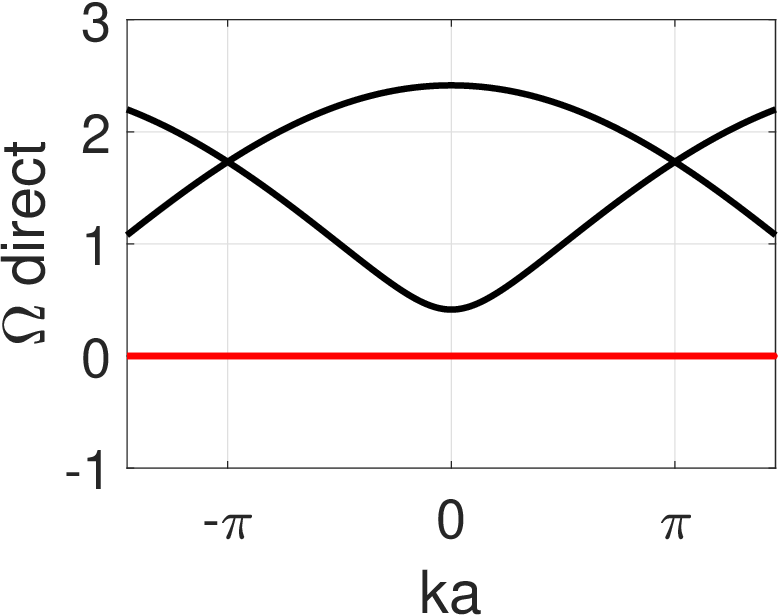}    &\includegraphics[height=\y cm, valign=c]{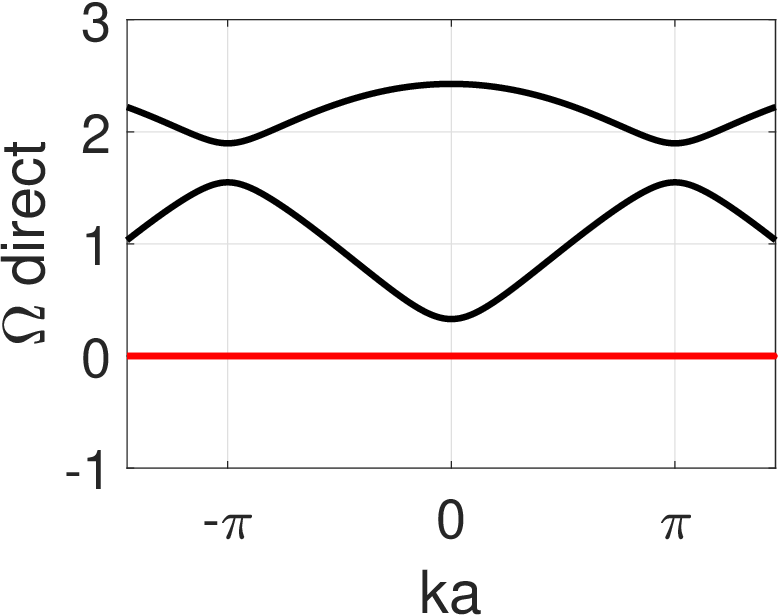} &  \includegraphics[height=\y cm, valign=c]{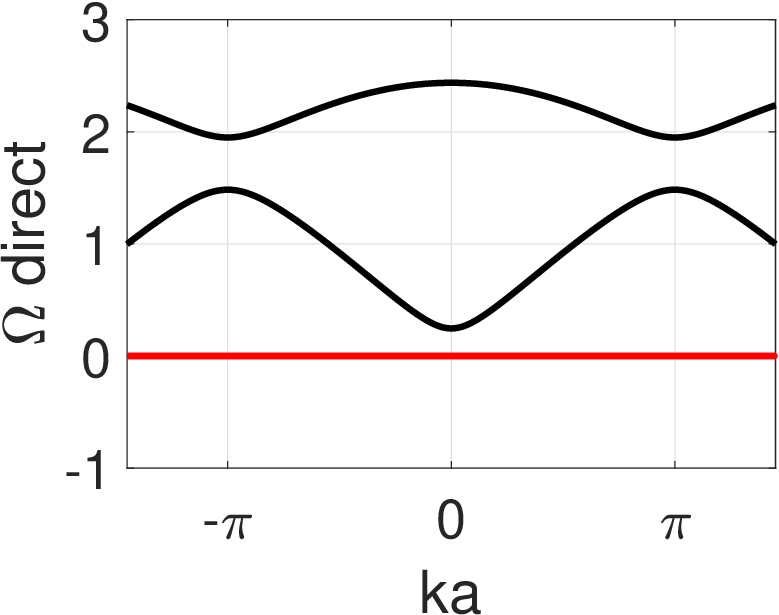} & \includegraphics[height=\y cm, valign=c]{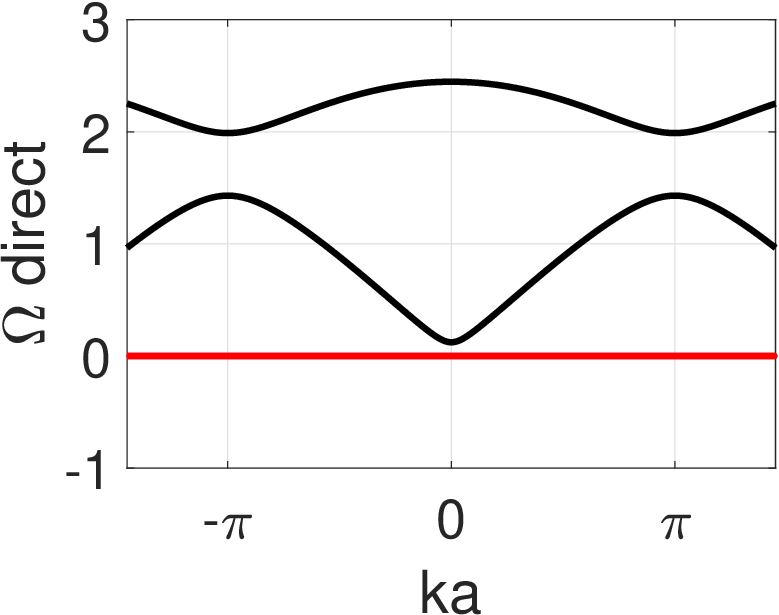}
 \\
 \includegraphics[height=\y cm, valign=c]{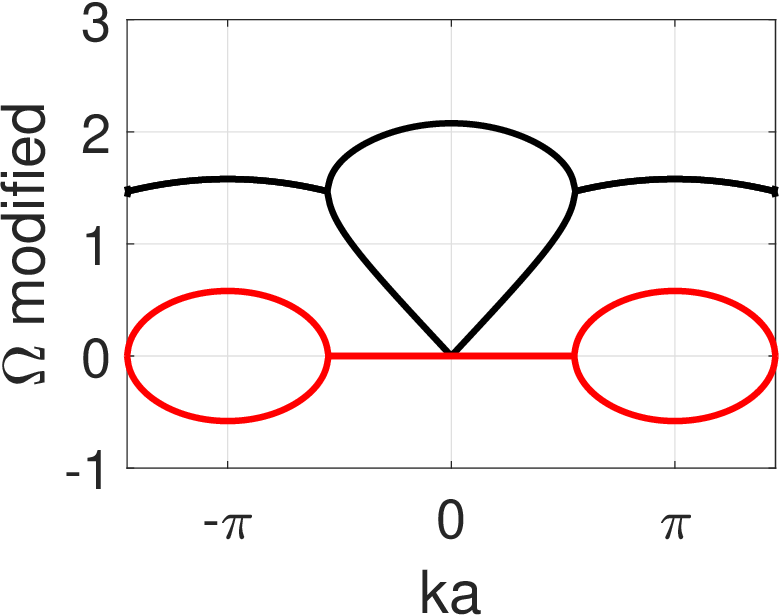}    & 
    \includegraphics[height=\y cm, valign=c]{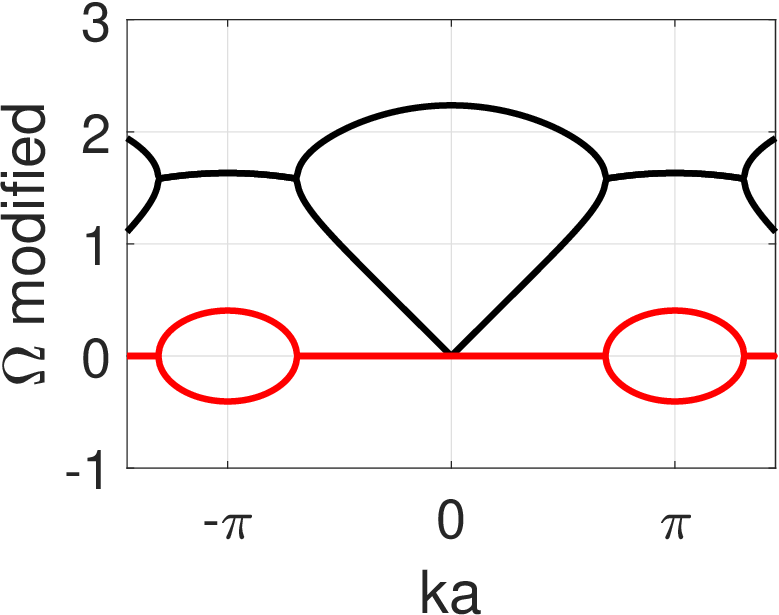}    &\includegraphics[height=\y cm, valign=c]{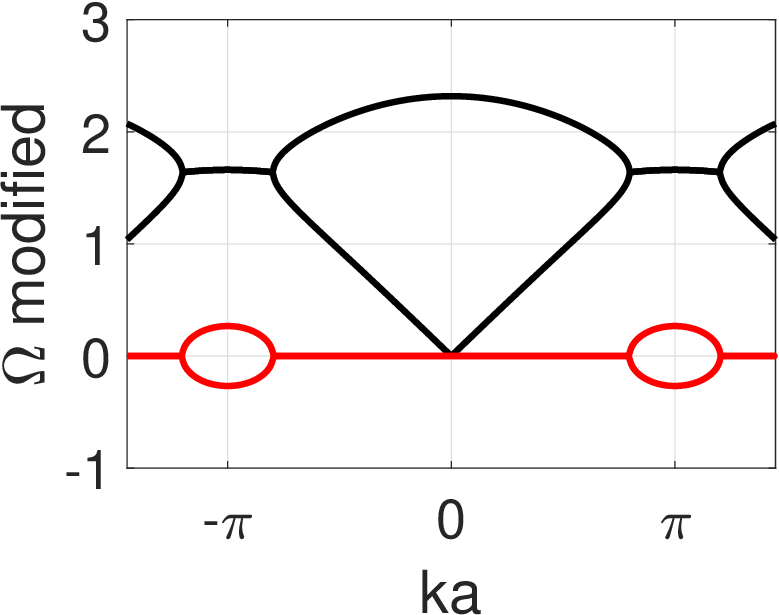} &  \includegraphics[height=\y cm, valign=c]{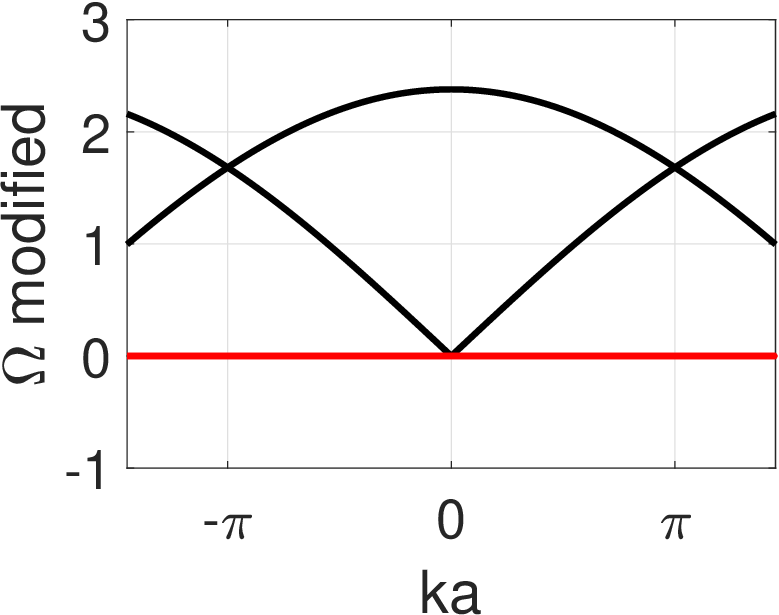} & \includegraphics[height=\y cm, valign=c]{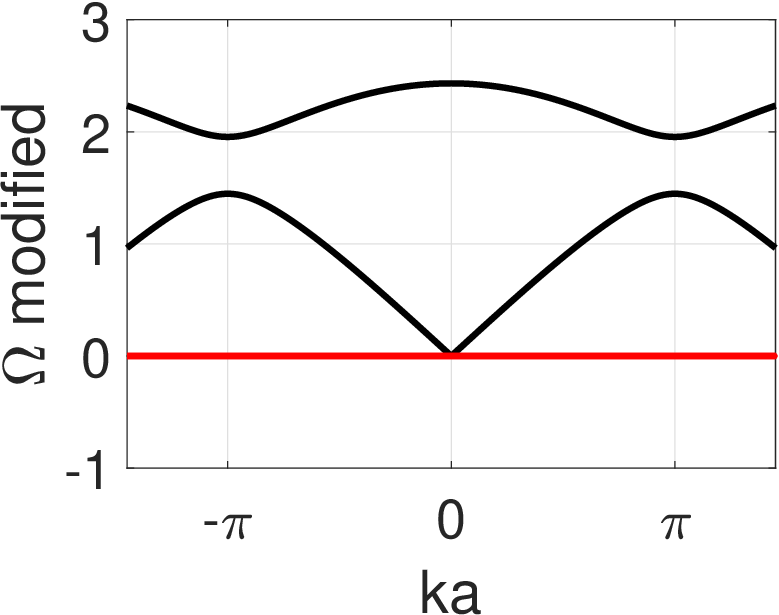}
 \\
 \includegraphics[height=\y cm, valign=c]{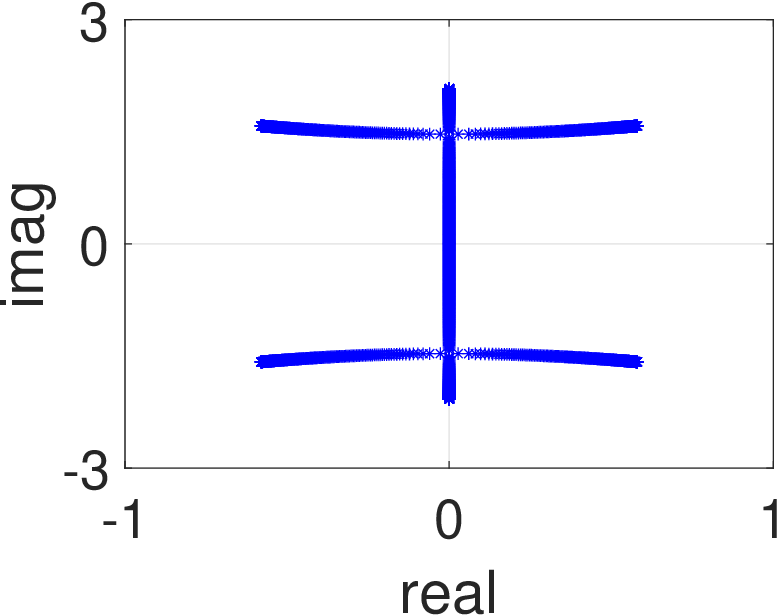}    & 
    \includegraphics[height=\y cm, valign=c]{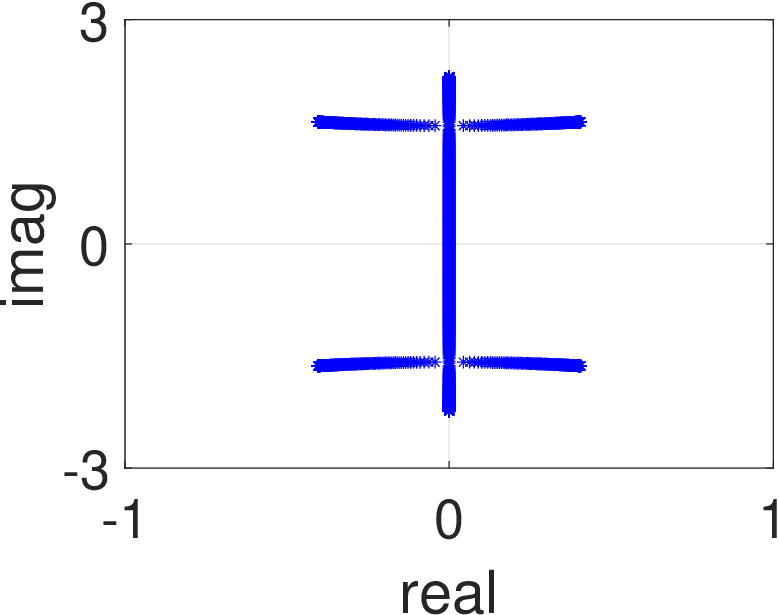}    &\includegraphics[height=\y cm, valign=c]{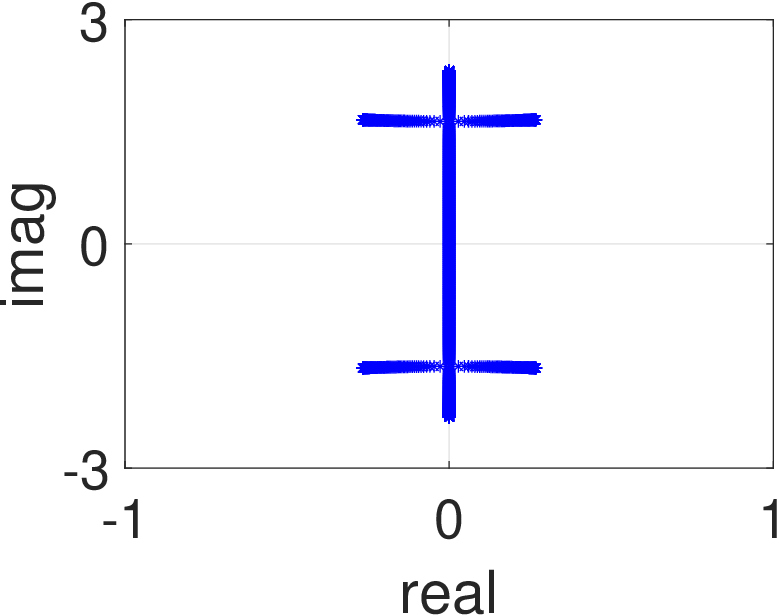} &  \includegraphics[height=\y cm, valign=c]{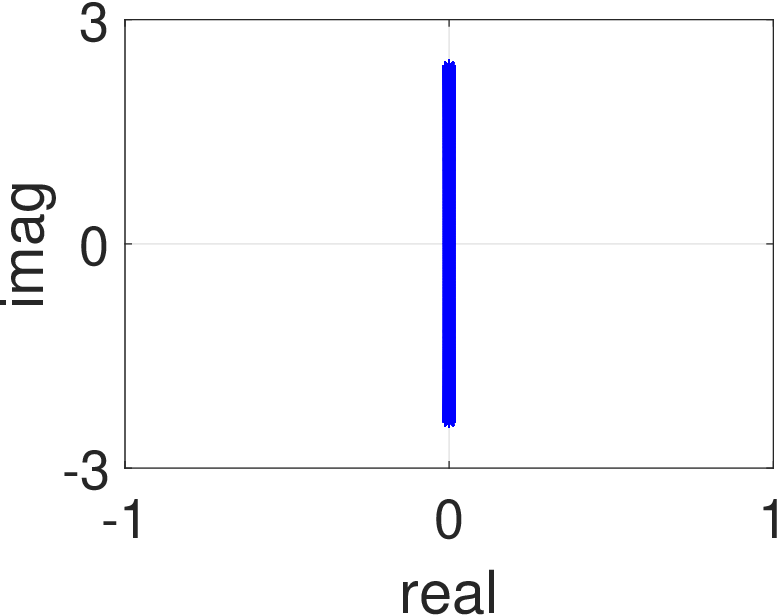} & \includegraphics[height=\y cm, valign=c]{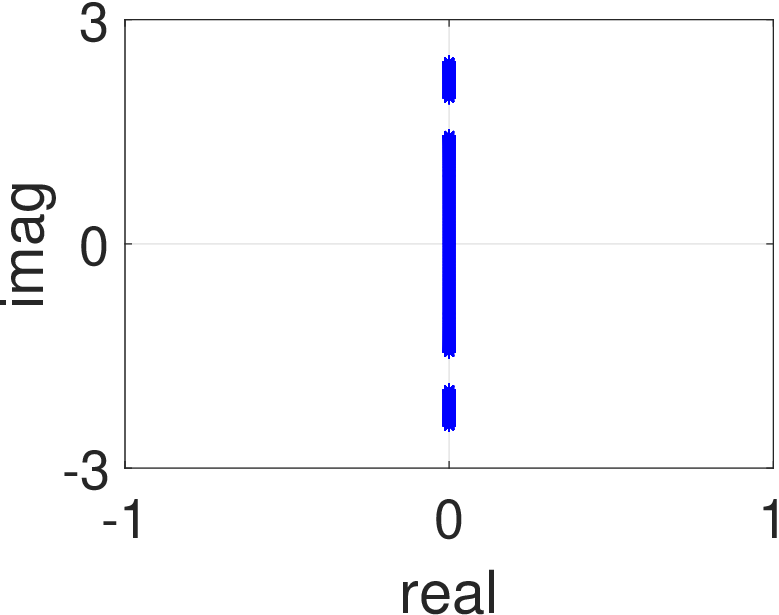}
\end{tabular} 
    \caption{\textbf{The model, stability and dispersion.} (a) The lattice schematic. (b) Stability diagram in the $\gamma-\eta$ plane distinguishing the actual stable region (dark gray) from the region aliasing as stable (light gray), using the modified mapping \eqref{eq:v_g_classical} (red curve). The direct mapping \eqref{eq:v_g_quantum} (black line) and the different working points are also plotted. (c)-(h) Dispersion of the quantum system, of the classical system for direct and modified mappings, and the system poles, for the square, bullet, triangle, diamond, circle, and star working points, respectively.}
    \label{fig:poles_direct}
\end{figure*}

Controlling wave propagation velocity in materials is of interest in diverse applications of Physics and Engineering. For example, the concept of anti-parallel group and phase velocities in metamaterials was studied to generate negative refraction \cite{pendry2004negative}, and the associated applications of sub-wavelength waveguides that overcome diffraction limits \cite{eleftheriades2005negative}, photon tunneling \cite{zhang2002unusual}, and more \cite{engheta2005positive}.
Phase velocities higher than the base value were invoked for electromagnetic and acoustic
cloaking \cite{pendry2006controlling,cummer2007one}, field focusing, and lensing \cite{kwon2010transformation}. 
Increasing the group velocity of a given medium or material is usually the goal in electromagnetic research due to the desire for superluminal propagation \cite{hrabar2013ultra,long2017observation,duggan2022stability} (within the fundamental limitations), with applications in communications \cite{zhang2011superluminal},
leaky-wave antennas \cite{sievenpiper2011superluminal} and broadband cloaking \cite{wong2017optical}. The leading approach in \cite{hrabar2013ultra,long2017observation,duggan2022stability} was incorporating active elements in a waveguide to create an effective medium with refractive index smaller than 1. 

Another approach to exceed the group velocity of a given medium or metamaterial remains within the lattice regime, i.e. without the requirement for a subwavelength spacing of the sites.
This is achieved by lifting up the crossing point of the dispersion cones, resulting in a steeper slope. 
In photonic lattices 
\cite{szameit2011p,ramezani2012exceptional}, for example, such a lifting was enabled by introducing gain and loss in the optical waveguides, balanced with an overall lattice stretching. 
Consequently, the wavepackets traveled with velocity proportional to the square root of the gain parameter.
The lattice was described by non-Hermitian  Hamiltonians \cite{berry2004physics,ashida2020non}, featuring complex-valued spectrum.
Here, our goal is to realize, via non-Hermiticity, an analogous control over group velocity in classical lattices.

Utilizing non-Hermitian systems for advanced wave dynamics came recently into spotlight with unidirectional invisibility, cloaking and focusing, coherent absorption, and more \cite{makris2008beam,zhu2014p,sounas2015unidirectional,fleury2015invisible,shi2016accessing,achilleos2017non,el2018non,gu2021controlling,gu2021acoustic,cao2022design, stojanoska2022non}, both on quantum and classical platforms.  
These applications, as well as the group velocity increase, require the underlying structural couplings to include either onsite gain, or some incompatibility with classical physics due to being complex-valued, directional, lacking restoring forces, etc. \cite{zhai2019active,sasmal2020broadband,rosa2020dynamics,geib2021tunable,raval2021experimental,baz2022breaking,jin2022non}.
Realization of such couplings requires external energy to be applied to the system, usually by embedding an active control mechanism in the host structure. This can be achieved either by direct feedback elements, such as operational amplifiers in electric circuit lattices \cite{hofmann2019chiral,helbig2020generalized,jana2023gravitational,jana2023tunneling,zhu2023higher,langfeldt2023controlling}, or by electronic controllers, which process real-time measurements of the dynamical response \cite{rosa2020dynamics,sirota2020non,zhang2021acoustic,wen2023acoustic}. 

The resulting systems, however, might feature a dynamical stability problem. This implies that instead of exhibiting the non-Hermitian phenomenon, the system undergoes a divergent time domain response
\cite{szameit2011p,rosa2020dynamics,duggan2022stability}.  
Here, we focus at designing the non-Hermitian systems in an inherently stable form, leading to bounded responses in time domain.
We begin with a simple model, a two-site periodic lattice of the constant $a$, as depicted in Fig. \ref{fig:poles_direct}(a). The sites $A$ and $B$ are distinguished by the local addition of gain and loss of magnitude $\gamma$, and by the near neighbor coupling $\eta$ and 1. 
In the quantum scenario, $\gamma$ represents on-site staggered potential, whereas $\eta$ is the inter-site electron hopping strength. The infinite lattice is governed by the non-Hermitian Bloch Hamiltonian $\mathcal{H}_q$,
\begin{equation}   \label{eq:Bloch_Hamiltonian}
    \mathcal{H}_q=\mathcal{H}_0-i\gamma\sigma_z, \; \mathcal{H}_0=\left(\begin{array}{cc} 0 & f \\ f* & 0 \end{array}\right), \; f=-\left[\eta+e^{-ika}\right],
\end{equation}
and $\sigma_z$ is Pauli matrix. The general energy spectrum of $\mathcal{H}_q$ is complex-valued. However, for a particular value $\gamma=\gamma^*$ the PT-symmetric phase is restored \cite{bender1998real}, yielding a pure real spectrum. We calculate the quantum group velocity $v_q$ at the vicinity of the exceptional points $ka=\pm\pi$, which are the crossing point of the bands.
$\gamma^*$ and $v_q$ are given by \cite{supplementary}
\begin{equation}  \label{eq:v_g_quantum}
    \gamma^*=\eta-1 \quad , \quad |v_q|\propto\sqrt{\gamma^*}
\end{equation}
cells per second. The atomic lattice, or equivalent photonic lattices in one and two dimensions \cite{szameit2011p,ramezani2012exceptional} (for which $\eta$ then represents stretching), were studied to support this type of control over the group velocity.
%
Here, our goal is to obtain an equivalent classical system in which the group velocity is controlled similarly to \eqref{eq:v_g_quantum}, theoretically leading to arbitrary fast propagating waves in the lattice. 
The quantum electron hopping is then translated into classical wave motion, such as displacement waves in mechanical mass-spring lattices, current waves in electrical transmission lines, sound waves in acoustic waveguides, etc. 
In any of the above classical platforms, the gray and the black couplings of Fig. \ref{fig:poles_direct}(a) indicate different compliance, $1$ and $\eta$, to energy transmission between each corresponding pair of sites, whereas the red and blue couplings respectively indicate onsite gain and loss.

Regardless of the platform, the underlying dynamics is of second order in time, and has the form $\Ddot{\textbf{Y}}+\omega_0K^v\dot{\textbf{Y}}+\omega_0^2K^y\textbf{Y}=\textbf{0}$. Here, $\textbf{Y}$ is the sites response vector, and $\omega_0$ is the characteristic frequency. $K^y$ and $K^v$ are the associated dynamical matrices, which respectively store the couplings between the lattice degrees of freedom, and between their rates.
Our task is to derive $K^y$ and $K^v$ that in the traveling harmonic wave regime $e^{i(kna-\omega t)}$ will yield an equivalent dynamics to the effective Hamiltonian in \eqref{eq:Bloch_Hamiltonian}, with the nondimensional frequency $\Omega=\omega/\omega_0$ replacing the energy spectrum. 
\begin{figure}[htpb]
    \centering
    \begin{tabular}{c}
    \textbf{(a)} \\
          \includegraphics[width=8.5 cm, valign=c]{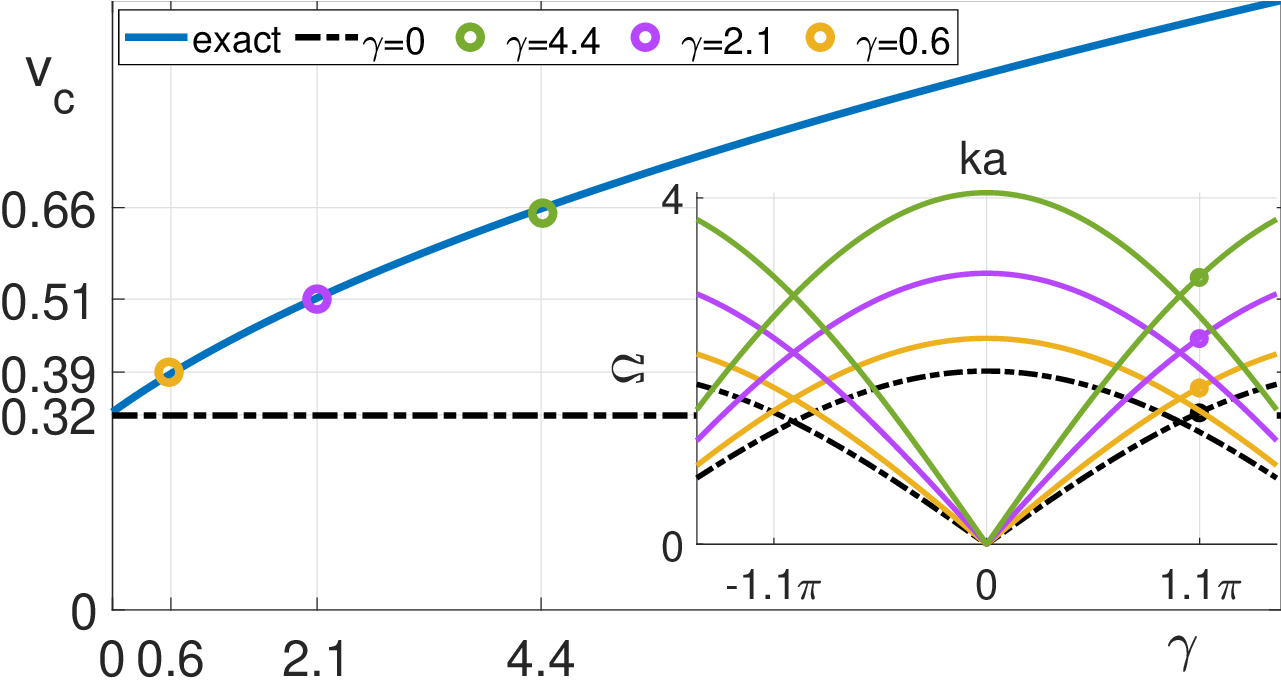}
     \end{tabular}
        \setlength{\tabcolsep}{0pt}
     \begin{tabular}{c c}
      \textbf{(b)} $\gamma=0$ ($\eta=1$)  &  \textbf{(c)} $\gamma=0.6$ ($\eta=2$)   \\
       \includegraphics[height=3.2 cm, valign=c]{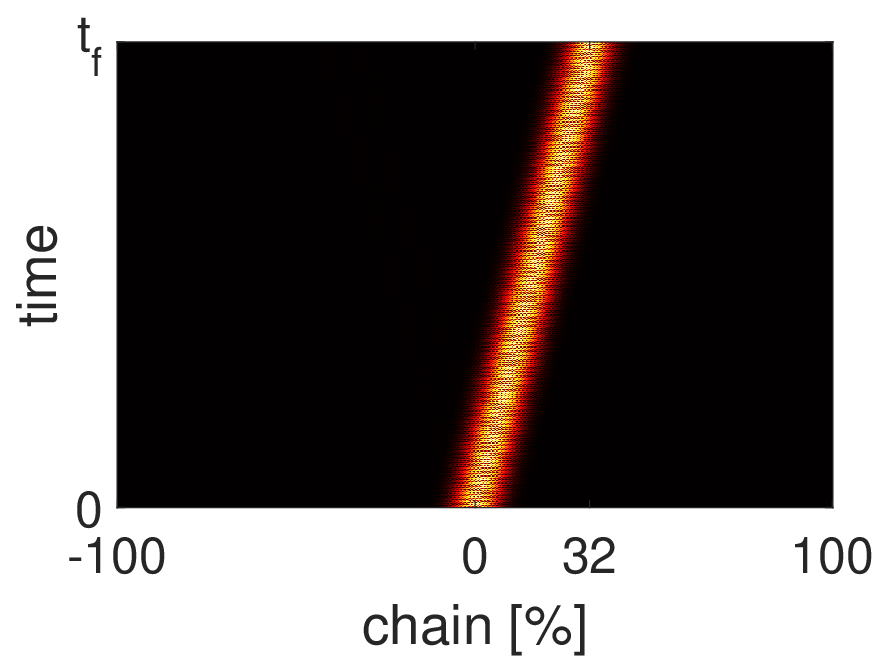}   & \includegraphics[height=3.2 cm, valign=c]{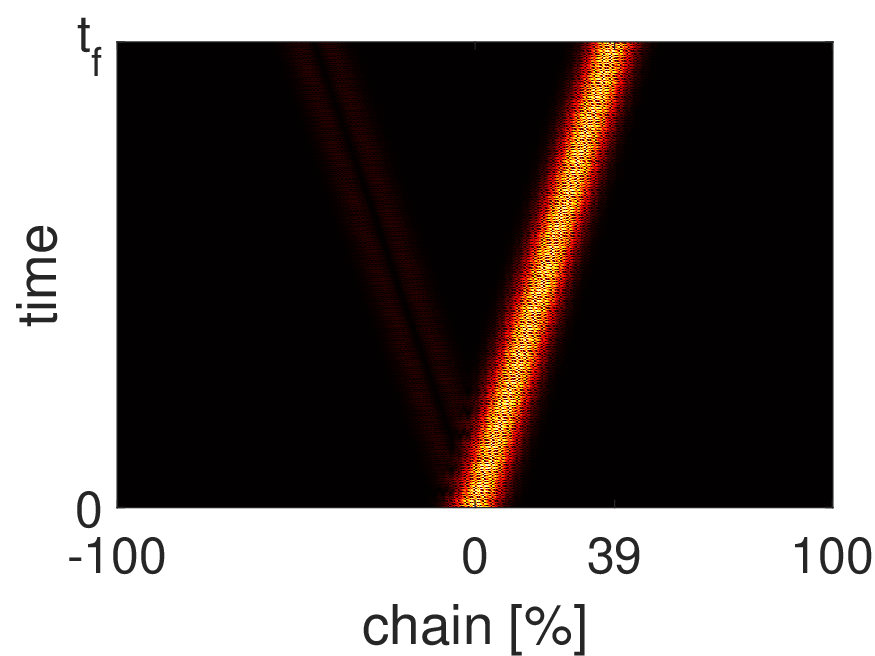} \\
       \textbf{(d)} $\gamma=2.1$ ($\eta=6$) &  \textbf{(e)} $\gamma=4.4$ ($\eta=17$) \\
       \includegraphics[height=3.2 cm, valign=c]{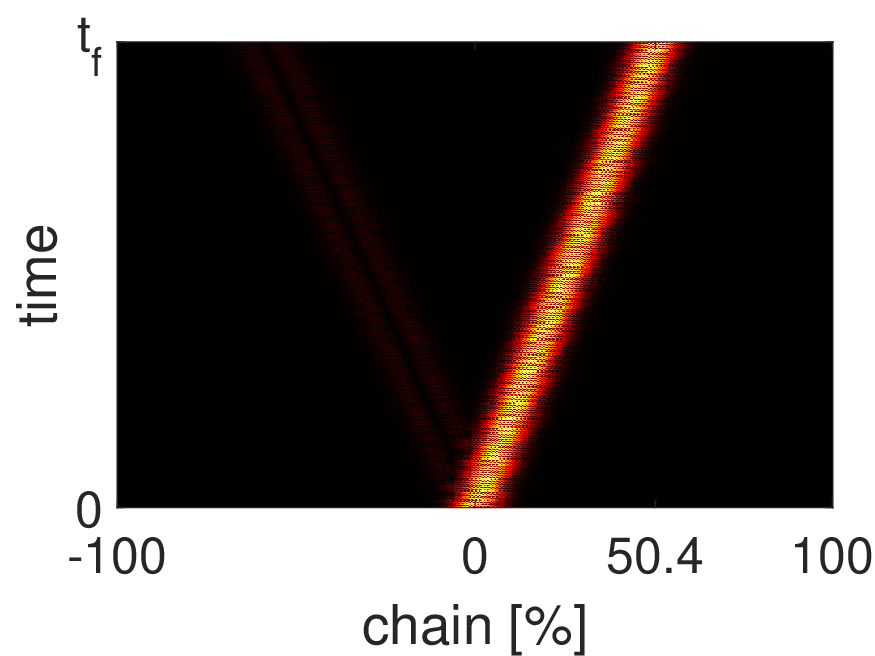}  & \includegraphics[height=3.2 cm, valign=c]{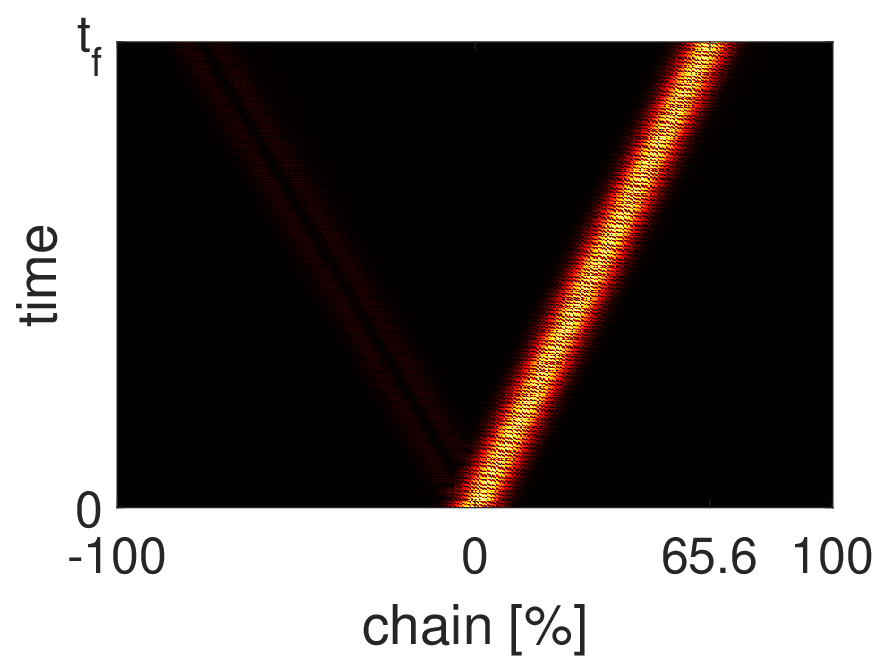}
     \end{tabular}   
     \centering
     \includegraphics[height=0.3 cm, valign=c]{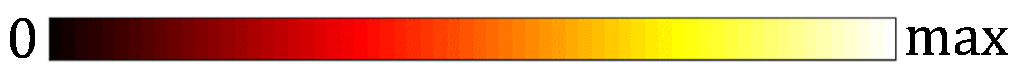}
    \caption{\textbf{Simulation of the modified classical model} \eqref{eq:n_dynamics}-\eqref{eq:v_g_classical}. (a) Inset: the excitation momentum $ka=3.46$ labeled on top of the frequency dispersion \eqref{eq:modified_mapping} for the Hermitian case $\gamma=0$ (black), and the non-Hermitian cases $\gamma=0.6$ (yellow), $\gamma=2.1$ (purple), and $\gamma=4.4$ (green). Main: the corresponding group velocity (dashed line and circles) calculated from the time domain simulations in (b)-(e) and plotted on top of the analytical result (blue curve) \eqref{eq:v_g_classical} as a function of $\gamma$. (b)-(e) Time domain responses of a generic normalized chain to an initial Bloch mode wavepacket for the different $\gamma$.}
    \label{fig:simulations}
\end{figure}
The common approach to back-translation of quantum Hamiltonians to the classical realm is adding a constant that shifts the real part of the frequency spectrum above the zero, which we denote by direct mapping. Here, this constant equals $1+\eta$, resulting in
\begin{equation}  \label{eq:Hamiltonian_c}
    \mathcal{H}_{cd}=(1+\eta)\textbf{I}+\mathcal{H}_q.
\end{equation}
The dynamical interpretation of the classical direct Hamiltonian $\mathcal{H}_{cd}$ in \eqref{eq:Hamiltonian_c} has two aspects. One is that the onsite restoring couplings, which are inherent to classical systems, are included in the equations of motion. 
Two is that in the time harmonic regime the gain and loss are represented by the steady state values of the rates $\pm\dot{\textbf{Y}}$, 
which are given by $i\textbf{Y}$ \cite{supplementary}.
Notably, the direct mapping \eqref{eq:Hamiltonian_c} implies the restoration of PT symmetry for $\gamma^*=\eta-1$, exactly as for the quantum system. 
It turns out, however, as we show next, that the mapping in \eqref{eq:Hamiltonian_c} might result in dynamical instability of the underlying classical system. 
We therefore suggest a modified mapping, in which 
the gain and loss are accounted for during the transient response as well, and not only in steady state. 
The $n_{th}$ unit cell time domain dynamics is then captured by
\begin{equation}  \label{eq:n_dynamics}
 K^y_n=\left[\arraycolsep=1.4pt\begin{array}{cccc}
    -1, & 1+\eta, & -\eta, & 0   \\
    0, & -\eta, & 1+\eta, & -1
  \end{array}\right], \; K^v_n=\left[\begin{array}{cc}
    -\gamma, & 0  \\
      0, & \gamma 
  \end{array}\right],
\end{equation}
where $K^y_n$ and $K^v_n$ are the $n_{th}$ blocks of the $K_y$ and $K_v$ matrices.
The resulting frequency spectrum is then obtained from
the classical modified Hamiltonian $\mathcal{H}_{cm}$ \cite{supplementary}
\begin{equation}  \label{eq:modified_mapping}
\mathcal{H}_{cm}=\left(\begin{array}{cc}
       \textbf{0}  & \textbf{I} \\
       \mathcal{H}_0+(1+\eta)\textbf{I}  & i\gamma\sigma_z
    \end{array}\right).
\end{equation}
The relation between $\gamma$ and $\eta$ that leads to restoration of PT symmetry, and the resulting classical group velocity $v_c$, read
\begin{equation}   \label{eq:v_g_classical}
    \gamma^*=\sqrt{2}\left(\sqrt{\eta}-1\right), \quad |v_c|=\frac{|\sin ka|\sqrt{\gamma+\sqrt{2}}}{4d(ka)\sqrt{\sqrt{2}\pm d(ka)}} 
\end{equation}
cells per second, where $d(ka)=\sqrt{1+\cos ka}$.
The relation between $\gamma$ and $\eta$ in \eqref{eq:v_g_classical} is no longer linear as in \eqref{eq:v_g_quantum}, but $v_c$ is proportional to $\sqrt{\gamma^*}$ similarly as $v_q$ up to scaling by a constant. 
The quantum and classical $\gamma^*$, \eqref{eq:v_g_quantum} and \eqref{eq:v_g_classical}, are depicted in Fig. \ref{fig:poles_direct}(b) by black and red curves, respectively, versus $\eta$. 
These curves delimit the regions of dynamical stability and instability, as we demonstrate in the following.
In Fig. \ref{fig:poles_direct}(c)-(h) we plot the quantum spectrum of \eqref{eq:Bloch_Hamiltonian}, and the classical spectra of \eqref{eq:Hamiltonian_c} and \eqref{eq:modified_mapping} for combinations of $\gamma$ and $\eta$ labeled by a square, a bullet, a triangle, a diamond, a circle, and a star in Fig. \ref{fig:poles_direct}(b). Alongside the different spectra, we plot the poles of the model in \eqref{eq:n_dynamics}, which are the eigenvalues of the matrix $\mathcal{A}=[0, \textbf{I} ; -K^v, -K^y]$ in the state space representation $\dot{\textbf{x}}=\mathcal{A}\textbf{x}$, where $\textbf{x}=[\textbf{Y} ; \dot{\textbf{Y}}]$. 

The square label in Fig. \ref{fig:poles_direct}(b) stands for $\eta=1$, $\gamma=0$, indicating the Hermitian case with no onsite gain and loss, and equal inter-site couplings. The quantum and classical spectra in this case are purely real and gap-less, as expected, with both direct and modified mappings appearing identical, and all the poles are on the imaginary axis indicating marginal stability, i.e. non-decaying and non-diverging harmonic wave propagation, Fig. \ref{fig:poles_direct}(c). 
For the bullet label we have $\eta=2$, $\gamma=1.3$, which is above all the PT symmetry relations. As a result, all the three spectra have an imaginary component, and the poles drift to the unstable right half plane region, Fig. \ref{fig:poles_direct} (d). 

The triangle label in Fig. \ref{fig:poles_direct}(e) indicates the PT symmetry relation of the quantum system \eqref{eq:v_g_quantum}, and of the classical system for the direct mapping. The imaginary component of the corresponding spectra annihilates, yielding the real bands touching at the exceptional point $ka=\pi$. However, contrary to the expectation for the direct mapping approach, the poles remain in the unstable region. This is clearly exhibited by the modified mapping, which includes imaginary spectrum. 
Further decreasing $\gamma$ to the diamond point opens a gap in the quantum and the direct classical spectra, Fig. \ref{fig:poles_direct}(f), which remain real, but the modified classical spectrum remains complex, and the system poles stay in the unstable region accordingly.

Only when decreasing toward the circle, the imaginary spectrum of the modified mapping disappears, and the poles entirely stick to the imaginary axis, Fig. \ref{fig:poles_direct}(g). 
For any $\gamma$ below the circle point, for example at the star point, a gap opens in the purely real spectrum, Fig. \ref{fig:poles_direct}(h), but the system is kept stable.
The relation in \eqref{eq:v_g_classical} therefore indicates the true restoration of the PT symmetric phase, and thus the true stability region, which is below the red curve in Fig. \ref{fig:poles_direct}(b). 
The region between the red and black curves thus indicates unstable dynamics, which aliases into a stable one when the direct mapping \eqref{eq:Hamiltonian_c} is used. The region above the black curve is unstable as well, appearing as such for all mappings.

\begin{figure*}[htpb]
    \centering
    \setlength{\tabcolsep}{-2pt}
\def\arraystretch{1.2} 
\begin{tabular}[t]{cc}
\begin{tabular}[t]{c}
\setlength{\tabcolsep}{0pt}
\begin{tabular}[t]{cc}
\textbf{(a)} & \textbf{(b)}  \\    
    \includegraphics[height=2.4 cm, valign=c]{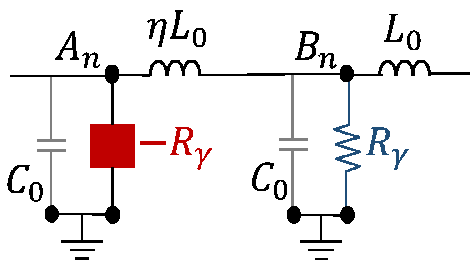} 
    &
    \includegraphics[height=3.7 cm, valign=c]{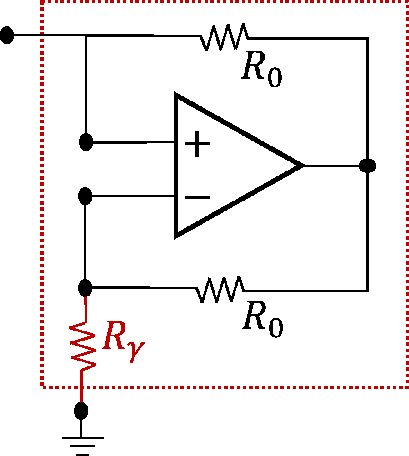}
\end{tabular} \\
    \def\arraystretch{1.4} 
    \begin{tabular}[t]{c}
         \textbf{(c)}  \\
         \includegraphics[height=2.6 cm, valign=t]{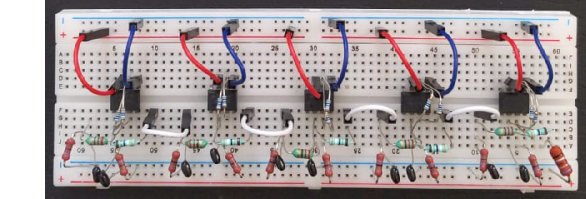} \\
          \textbf{(j)}  \\
         \includegraphics[height=4.0 cm, valign=t]{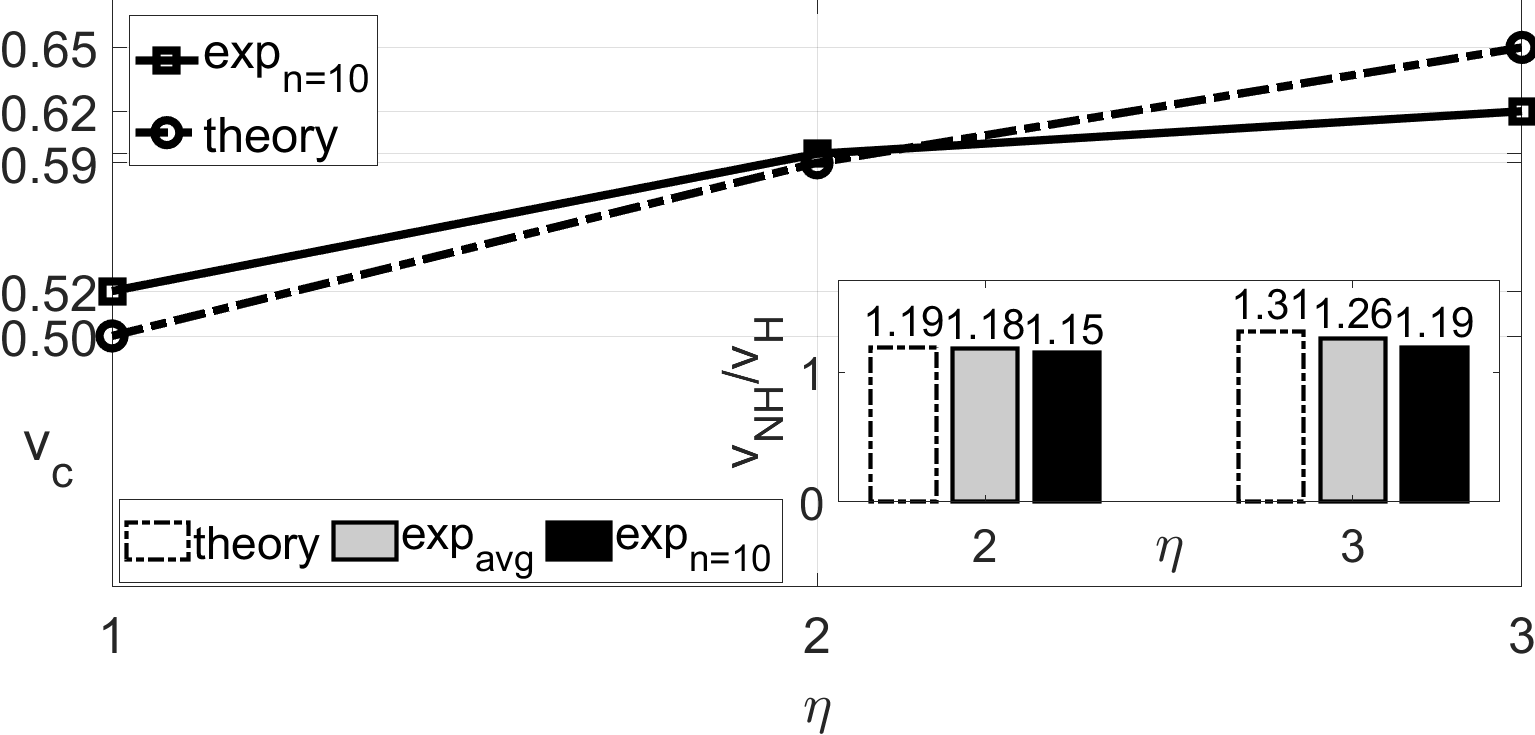}
    \end{tabular}
    \end{tabular}
    \begin{tabular}[t]{c}
\setlength{\tabcolsep}{2pt}
\def\arraystretch{1.3} 
\begin{tabular}[t]{cc}
    \textbf{(d)} $\eta=2$ Experiment & \textbf{(e)} $\eta=3$ Experiment \\   
    \includegraphics[height=3.8 cm, valign=t]{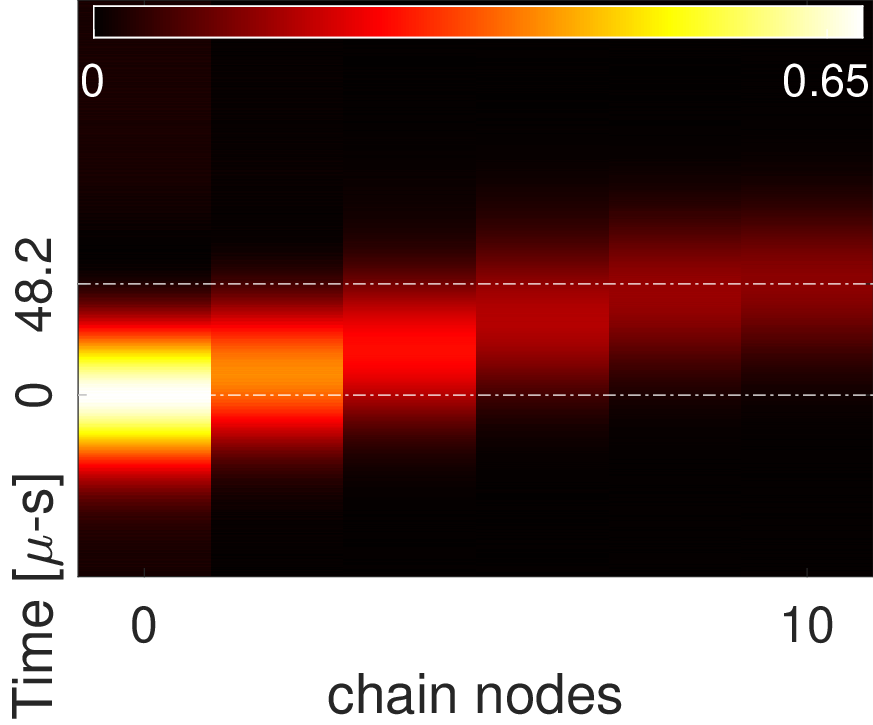} 
    & 
    \includegraphics[height=3.8 cm, valign=t]{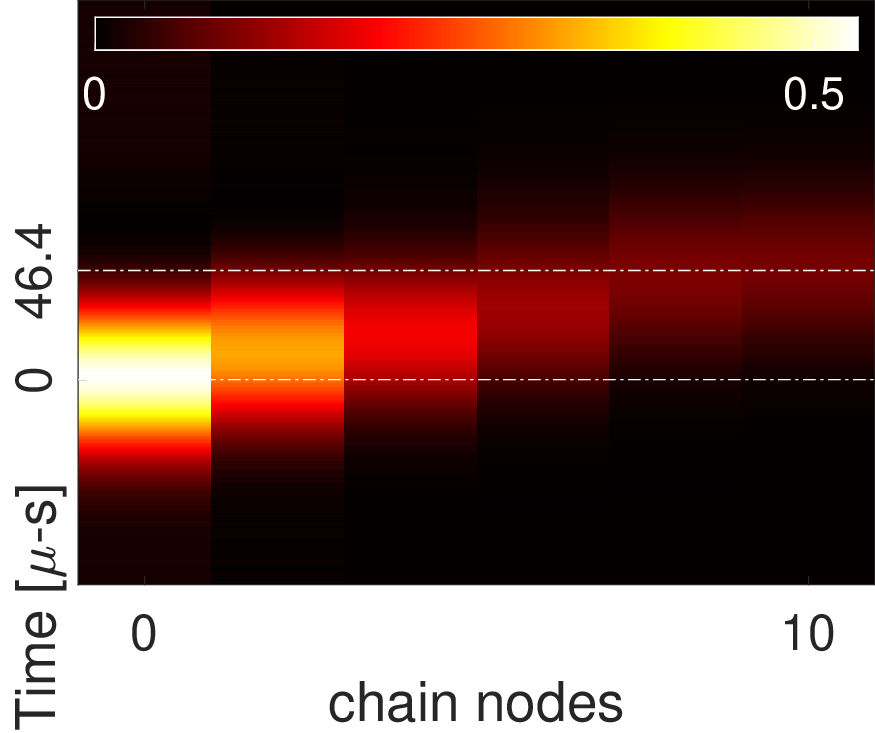} \end{tabular} \\
    \setlength{\tabcolsep}{0pt}
\def\arraystretch{1.3} 
\begin{tabular}[t]{cc}
\textbf{(f)} Experiment & \textbf{(g)} Experiment \\    
    \includegraphics[height=3.4 cm, valign=t]{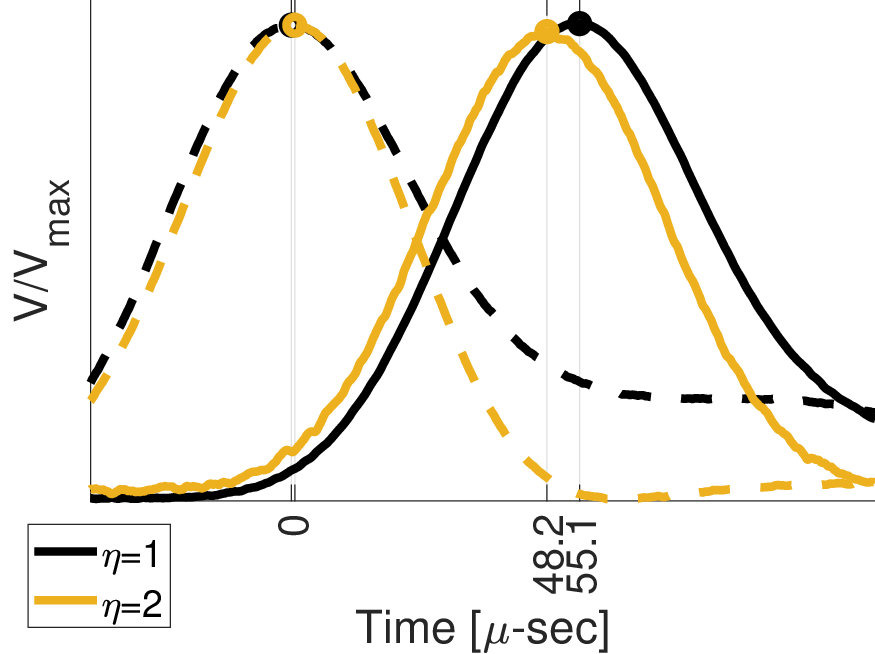} 
    & 
    \includegraphics[height=3.4 cm, valign=t]{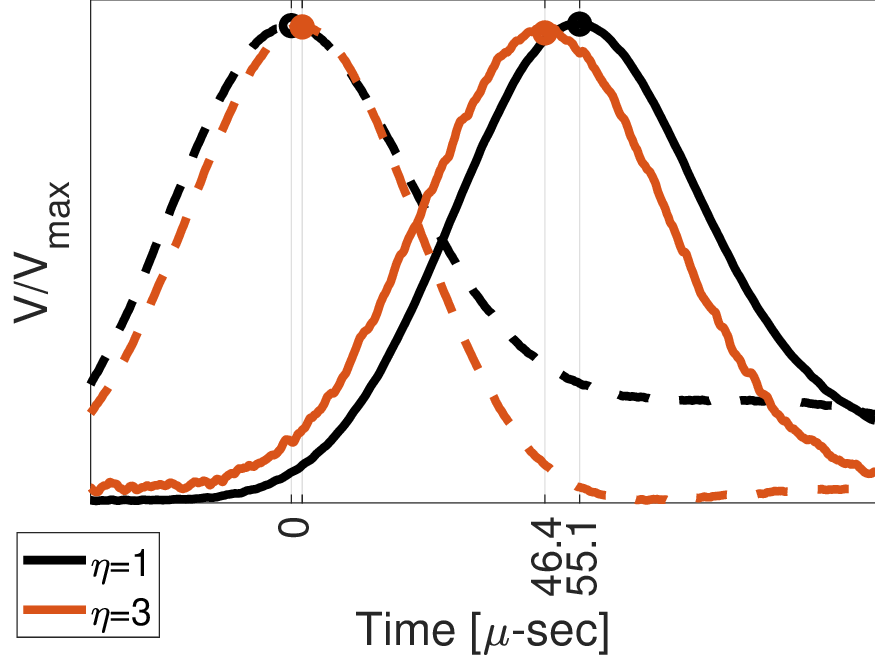} \\
    \textbf{(h)} Theory & \textbf{(i)} Theory \\    
    \includegraphics[height=3.4 cm, valign=t]{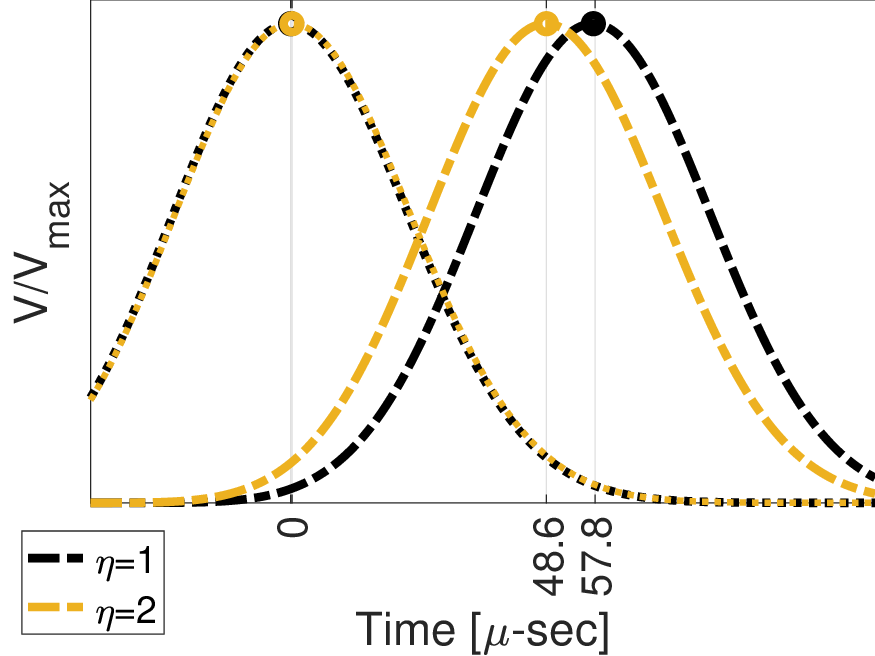} 
    & 
    \includegraphics[height=3.4 cm, valign=t]{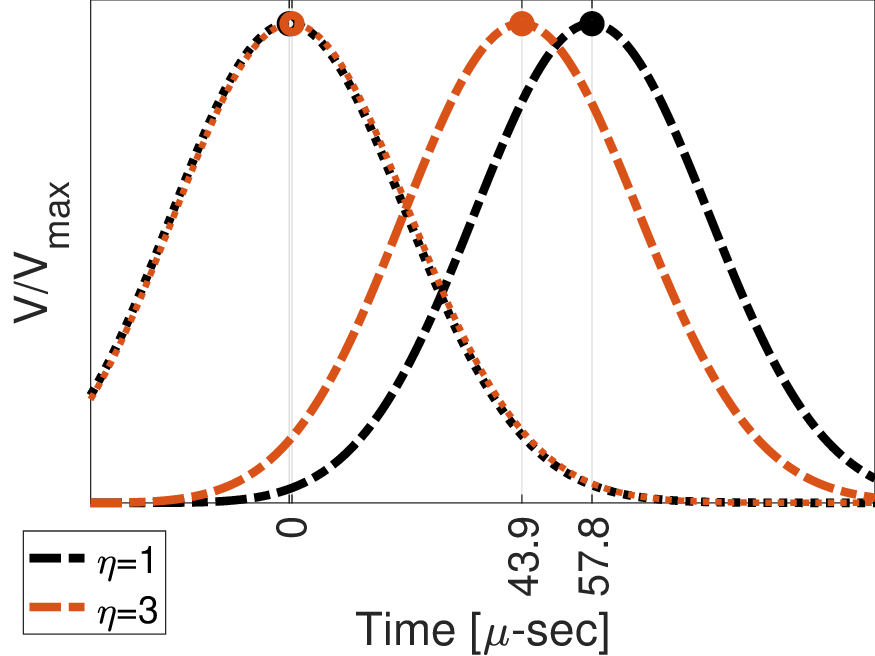}
\end{tabular}
\end{tabular}
\end{tabular}
    \caption{\textbf{Experimental demonstration of the fast wavepackets in a topoelectric metamaterial.} (a) The active element positioning in the $n_{th}$ unit cell of the nominal transmission-line, indicating gain (red) and loss (blue). (b) Detailed schematic of the active element, featuring an operational amplifier in a negative impedance converting setup, producing the required gain. (c) The experimental setup, featuring a 10-site chain. The Hermitian basis is comprised of ten RF inductors $L_0=220$ muH and ten ceramic capacitors $C_0=150$ nF. The non-Hermitian addition comprises five active and five passive cells. Each active cell includes an operational amplifier powered by 12 V DC voltage, two resistors $R_0=1$ kOhm, and one $R_\gamma=68$ or $39$ Ohm resistor, corresponding to $\eta=2$ or $3$, respectively. Each passive cell includes one $R_\gamma=68$ or $39$ Ohm resistor, accordingly. (d), (e) Measured voltage in time at all even nodes for $\eta=2,3$. Marked are the times of node 0 and 10. (f), (g) The measurements at node 0, dashed, and 10, solid, as a function of time for $\eta=2$, yellow, and $\eta=3$, orange, compared to the measurement for $\eta=1$, black. Marked are the arrival times. (h), (i) The theoretical results corresponding to (f), (g), with dashed-dotted for node 10 and dotted for node 0. (j) Main: the calculated group velocities in cells per second for $\eta=1,2$ and 3, for the measured results in (f), (g), solid, and the simulated results in (h), (i), dashed-dotted. Inset: the ratio of the non-Hermitian and the Hermitian system $v_c$ for $\eta=2$ and 3. Dashed-dotted white: theoretical. Solid grey: measured and averaged across all nodes. Solid black: measured at node 10.}
    \label{fig:experiment}
\end{figure*}

To illustrate the dynamics of the model in \eqref{eq:n_dynamics}-\eqref{eq:v_g_classical}, we first carry out a numerical experiment of a generic classical chain with $\gamma$ switching between $0$, $0.6$, $2.1$, and $4.4$. At all the cases we excite the chain center by the same initial wavepacket of momentum $ka=3.46$, as illustrated in Fig. \ref{fig:simulations}(a)-inset by black, yellow, purple, and green circles on top of the corresponding dispersion curves. Constructing the wavepacket from left Bloch eigenvectors for initial $\textbf{Y}$ and $\dot{\textbf{Y}}$, the bi-orthogonality of the non-Hermitian modes induces unidirectional propagation \cite{supplementary}, as depicted in Fig. \ref{fig:simulations}(b)-(e). For the four cases of $\gamma$ we stop the simulation at the same time instant $\mathrm{t_f}$, and the wavepacket reaches a farther site of the chain. 
This demonstrates that the velocity was indeed increased with $\gamma$.

The group velocity calculated from these simulations, in units of cells per second, is plotted in Fig. \ref{fig:simulations}(a)-main panel by yellow, purple, and green circles for the non-Hermitian cases $\gamma=0.6$, $2.2$, and $4.4$ on top of the analytical group velocity curve, blue, calculated from \eqref{eq:v_g_classical}. The Hermitian group velocity corresponding to $\gamma=0$ is plotted by a dashed black line. The simulated wavepacket velocities fit well the predicted theoretical square root dependency, accordingly growing as a function of the gain parameter $\gamma$, and for any $\gamma>0$ are higher than the underlying Hermitian case.

We now demonstrate this result in an electric circuit lattice, also known as topoelectrical metamaterial, in which the sites, the squares in Fig. \ref{fig:poles_direct}(a), are realized by capacitors $C_0$, and the couplings between the sites, the bars in Fig. \ref{fig:poles_direct}(a), are realized by inductors $L_0$. 
This capacitor-inductor transmission line, defined by the characteristic frequency $\omega_0=1/\sqrt{L_0C_0}$, constitutes the Hermitian basis of the lattice. 
The realization of the non-Hermitian part is illustrated in Fig. \ref{fig:experiment}(a). In the $n_{th}$ unit cell, the loss at the $B$ site is realized by a resistor $R_{\gamma}=\gamma\sqrt{C_0/L_0}$. 
To induce a corresponding gain at the $A$ site, we implement negative resistance of $-R_{\gamma}$. This is achieved using an active system captured by a red square, which is connected in parallel to the capacitor. 
It's key element is operational amplifier, Fig. \ref{fig:experiment}(b), which is supplied by constant voltage, and stands for both actuation and sensing. Such a system constitutes a negative impedance converter, featuring both forward and backward current flow through identical resistors $R_0$.
The backward flow is fed to the transmission line through a $R_{\gamma}$ resistor, yielding negative effective resistance $-R_{\gamma}$ between the terminals.
The system is then governed by \eqref{eq:n_dynamics}, with the onsite gain and loss $\pm\gamma$ \cite{supplementary}.

The experiment was carried out using the platform depicted in Fig. \ref{fig:experiment}(c). It comprised 10 $L_0$-$C_0$ junctions, i.e. 5 $A-B$ sites. 
We excited the lattice to the left of the first $A$ site, denoted by node 0, by a $2$ kHz voltage signal ($\Omega\approx 0.012$) of a 1 V amplitude, modulated by a Gaussian envelope, and performed three sets of measurements. The first set was for the Hermitian case $\eta=1$ and $\gamma=0$. In the other two sets the $\eta$ values were given by $2$ and $3$ ($ka\leq 0.025$), respectively realized by two and three $L_0$ inductors connected in parallel. The resulting values of $\gamma=0.6$ and $1.0$ were calculated from \eqref{eq:v_g_classical}. 
The voltage responses in time at nodes 0,2,4,6,8,10 are plotted in the heatmaps of Fig. \ref{fig:experiment}(d), (e), for $\eta=2$ and 3. An amplitude decay is observed, as a result of the circuit elements dissipation. The relative arrival times from node 0 to 10 are marked on top of the maps. 
The measured time responses at nodes 0 and 10 are depicted in Fig. \ref{fig:experiment}(f), (g), for $\eta=2$ and 3 on top of $\eta=1$. The corresponding theoretical responses are given in Fig. \ref{fig:experiment}(h), (i). 

The group velocities were then calculated for $\eta=1,2,3$ from the measurements according to $v_c=10/t_{10}/\omega_0$ for lattice constant 1, where $t_{10}$ denotes the arrival times, and depicted in Fig. \ref{fig:experiment}(j), main panel. 
It can be observed that the higher is $\eta$ (and thus $\gamma$), the higher is the group velocity, and the measured result follows quite well the theoretical growth trend. The increase that is obtained due to non-Hermiticity is demonstrated in the inset, in terms of the ratio $v_{\eta=2,3}$ over $v_{\eta=1}$. Three bars are presented: the theoretical, the experimental averaged over all the nodes, and the experimental for node 10. The latter two differ due to deviation in the circuit elements values. The best fitting was obtained for $\eta=2$ with $18\%$ and $15\%$ (average and node 10) velocity increase, compared to the theoretical $19\%$. For $\eta=3$ we obtained $26\%$ and $19\%$ of increase, compared to the theoretical $31\%$.

To summarize, in this work we proposed a method to increase the group velocity in non-Hermitian lattices, Fig. \ref{fig:poles_direct}(a), in a controlled manner via two design parameters, onsite gain and loss $\gamma$, and additional inter-site reactive coupling $\eta$. We derived a balancing nonlinear relation between $\gamma$ and $\eta$, Eq. \eqref{eq:v_g_classical}, which brought the velocity increase to be stable in a second order classical dynamics realm, Fig, \ref{fig:poles_direct}(b). The resulting group velocity featured a square root dependence on the $\gamma$ parameter, similarly to an analogous quantum system, although the quantum balancing $\gamma$-$\eta$ relation, Eq. \eqref{eq:v_g_quantum}, was linear. 
We demonstrated the velocity increase in simulations, Fig. \ref{fig:simulations}, and in an experiment, Fig. \ref{fig:experiment}. In the latter, we tested a 10-cell electrical transmission line with $\gamma=0.6$ and $1.0$, 
in which the gain was generated by operational amplifiers in a feedback setup.
Our design enables a stable fast wavepacket propagation for any frequency within the lattice passband.

\section*{Acknowledgements}

\textit{This research was supported in part by the Israel Science Foundation Grants No. 2177/23 and 2876/23. 
The authors are grateful to Arkadi Rafalovich for his invaluable help with the technical aspects of the experimental setup.}

\bibliographystyle{IEEEtran}

\bibliography{paper}

\begin{thebibliography}{10}
\providecommand{\url}[1]{#1}
\csname url@samestyle\endcsname
\providecommand{\newblock}{\relax}
\providecommand{\bibinfo}[2]{#2}
\providecommand{\BIBentrySTDinterwordspacing}{\spaceskip=0pt\relax}
\providecommand{\BIBentryALTinterwordstretchfactor}{4}
\providecommand{\BIBentryALTinterwordspacing}{\spaceskip=\fontdimen2\font plus
\BIBentryALTinterwordstretchfactor\fontdimen3\font minus \fontdimen4\font\relax}
\providecommand{\BIBforeignlanguage}[2]{{%
\expandafter\ifx\csname l@#1\endcsname\relax
\typeout{** WARNING: IEEEtran.bst: No hyphenation pattern has been}%
\typeout{** loaded for the language `#1'. Using the pattern for}%
\typeout{** the default language instead.}%
\else
\language=\csname l@#1\endcsname
\fi
#2}}
\providecommand{\BIBdecl}{\relax}
\BIBdecl

\bibitem{pendry2004negative}
J.~B. Pendry, ``Negative refraction,'' \emph{{C}ontemporary {P}hysics}, vol.~45, no.~3, pp. 191--202, 2004.

\bibitem{eleftheriades2005negative}
G.~V. Eleftheriades and K.~G. Balmain, \emph{Negative-refraction metamaterials: fundamental principles and applications}.\hskip 1em plus 0.5em minus 0.4em\relax John Wiley \& Sons, 2005.

\bibitem{zhang2002unusual}
Z.~Zhang and C.~Fu, ``Unusual photon tunneling in the presence of a layer with a negative refractive index,'' \emph{Applied {P}hysics {L}etters}, vol.~80, no.~6, pp. 1097--1099, 2002.

\bibitem{engheta2005positive}
N.~Engheta and R.~W. Ziolkowski, ``A positive future for double-negative metamaterials,'' \emph{{IEEE} {T}ransactions on {M}icrowave {T}heory and {T}echniques}, vol.~53, no.~4, pp. 1535--1556, 2005.

\bibitem{pendry2006controlling}
J.~B. Pendry, D.~Schurig, and D.~R. Smith, ``Controlling electromagnetic fields,'' \emph{Science}, vol. 312, no. 5781, pp. 1780--1782, 2006.

\bibitem{cummer2007one}
S.~A. Cummer and D.~Schurig, ``One path to acoustic cloaking,'' \emph{New {J}ournal of {P}hysics}, vol.~9, no.~3, p.~45, 2007.

\bibitem{kwon2010transformation}
D.-H. Kwon and D.~H. Werner, ``Transformation electromagnetics: An overview of the theory and applications,'' \emph{IEEE Antennas and Propagation Magazine}, vol.~52, no.~1, pp. 24--46, 2010.

\bibitem{hrabar2013ultra}
S.~Hrabar, I.~Krois, I.~Bonic, and A.~Kiricenko, ``Ultra-broadband simultaneous superluminal phase and group velocities in non-{F}oster epsilon-near-zero metamaterial,'' \emph{Applied Physics Letters}, vol. 102, no.~5, 2013.

\bibitem{long2017observation}
J.~Long and D.~F. Sievenpiper, ``The observation of dispersionless superluminal propagation in a non-{F}oster loaded waveguide and its fundamental limitations,'' \emph{IEEE Transactions on Microwave Theory and Techniques}, vol.~66, no.~2, pp. 762--773, 2017.

\bibitem{duggan2022stability}
R.~Duggan, H.~Moussa, Y.~Ra’di, D.~L. Sounas, and A.~Al{\`u}, ``Stability bounds on superluminal propagation in active structures,'' \emph{Nature Communications}, vol.~13, no.~1, p. 1115, 2022.

\bibitem{zhang2011superluminal}
L.~Zhang, L.~Zhan, K.~Qian, J.~Liu, Q.~Shen, X.~Hu, S.~Luo \emph{et~al.}, ``Superluminal propagation at negative group velocity in optical fibers based on {B}rillouin lasing oscillation,'' \emph{Physical Review Letters}, vol. 107, no.~9, p. 093903, 2011.

\bibitem{sievenpiper2011superluminal}
D.~F. Sievenpiper, ``Superluminal waveguides based on non-{F}oster circuits for broadband leaky-wave antennas,'' \emph{IEEE Antennas and Wireless Propagation Letters}, vol.~10, pp. 231--234, 2011.

\bibitem{wong2017optical}
Z.~J. Wong, Y.~Wang, K.~O’Brien, J.~Rho, X.~Yin, S.~Zhang, N.~Fang, T.-J. Yen, and X.~Zhang, ``Optical and acoustic metamaterials: superlens, negative refractive index and invisibility cloak,'' \emph{Journal of Optics}, vol.~19, no.~8, p. 084007, 2017.

\bibitem{szameit2011p}
A.~Szameit, M.~C. Rechtsman, O.~Bahat-Treidel, and M.~Segev, ``{PT}-symmetry in honeycomb photonic lattices,'' \emph{Physical Review A}, vol.~84, no.~2, p. 021806, 2011.

\bibitem{ramezani2012exceptional}
H.~Ramezani, T.~Kottos, V.~Kovanis, and D.~N. Christodoulides, ``Exceptional-point dynamics in photonic honeycomb lattices with {PT} symmetry,'' \emph{Physical Review A}, vol.~85, no.~1, p. 013818, 2012.

\bibitem{berry2004physics}
M.~V. Berry, ``Physics of non {H}ermitian degeneracies,'' \emph{Czechoslovak Journal of Physics}, vol.~54, no.~10, pp. 1039--1047, 2004.

\bibitem{ashida2020non}
Y.~Ashida, Z.~Gong, and M.~Ueda, ``Non-{H}ermitian physics,'' \emph{Advances in Physics}, vol.~69, no.~3, pp. 249--435, 2020.

\bibitem{makris2008beam}
K.~G. Makris, R.~El-Ganainy, D.~Christodoulides, and Z.~H. Musslimani, ``Beam dynamics in {PT} symmetric optical lattices,'' \emph{Physical Review Letters}, vol. 100, no.~10, p. 103904, 2008.

\bibitem{zhu2014p}
X.~Zhu, H.~Ramezani, C.~Shi, J.~Zhu, and X.~Zhang, ``{PT}-symmetric acoustics,'' \emph{Physical Review X}, vol.~4, no.~3, p. 031042, 2014.

\bibitem{sounas2015unidirectional}
D.~L. Sounas, R.~Fleury, and A.~Al{\`u}, ``Unidirectional cloaking based on metasurfaces with balanced loss and gain,'' \emph{Physical Review Applied}, vol.~4, no.~1, p. 014005, 2015.

\bibitem{fleury2015invisible}
R.~Fleury, D.~Sounas, and A.~Alu, ``An invisible acoustic sensor based on parity-time symmetry,'' \emph{Nature Communications}, vol.~6, no.~1, pp. 1--7, 2015.

\bibitem{shi2016accessing}
C.~Shi, M.~Dubois, Y.~Chen, L.~Cheng, H.~Ramezani, Y.~Wang, and X.~Zhang, ``Accessing the exceptional points of parity-time symmetric acoustics,'' \emph{Nature Communications}, vol.~7, no.~1, pp. 1--5, 2016.

\bibitem{achilleos2017non}
V.~Achilleos, G.~Theocharis, O.~Richoux, and V.~Pagneux, ``Non-{H}ermitian acoustic metamaterials: {R}ole of exceptional points in sound absorption,'' \emph{Physical {R}eview {B}}, vol.~95, no.~14, p. 144303, 2017.

\bibitem{el2018non}
R.~El-Ganainy, K.~G. Makris, M.~Khajavikhan, Z.~H. Musslimani, S.~Rotter, and D.~N. Christodoulides, ``Non-{H}ermitian physics and {PT} symmetry,'' \emph{Nature Physics}, vol.~14, no.~1, pp. 11--19, 2018.

\bibitem{gu2021controlling}
Z.~Gu, H.~Gao, P.-C. Cao, T.~Liu, X.-F. Zhu, and J.~Zhu, ``Controlling sound in non-{H}ermitian acoustic systems,'' \emph{Physical Review Applied}, vol.~16, no.~5, p. 057001, 2021.

\bibitem{gu2021acoustic}
Z.~Gu, T.~Liu, H.~Gao, S.~Liang, S.~An, and J.~Zhu, ``Acoustic coherent perfect absorber and laser modes via the non-{H}ermitian dopant in the zero index metamaterials,'' \emph{Journal of Applied Physics}, vol. 129, no.~23, p. 234901, 2021.

\bibitem{cao2022design}
L.~Cao, Y.~Zhu, S.~Wan, Y.~Zeng, and B.~Assouar, ``On the design of non-{H}ermitian elastic metamaterial for broadband perfect absorbers,'' \emph{International Journal of Engineering Science}, vol. 181, p. 103768, 2022.

\bibitem{stojanoska2022non}
K.~Stojanoska and C.~Shen, ``Non-{H}ermitian planar elastic metasurface for unidirectional focusing of flexural waves,'' \emph{Applied Physics Letters}, vol. 120, no.~24, 2022.

\bibitem{zhai2019active}
Y.~Zhai, H.-S. Kwon, and B.-I. Popa, ``Active willis metamaterials for ultracompact nonreciprocal linear acoustic devices,'' \emph{Physical Review B}, vol.~99, no.~22, p. 220301, 2019.

\bibitem{sasmal2020broadband}
A.~Sasmal, N.~Geib, B.-I. Popa, and K.~Grosh, ``Broadband nonreciprocal linear acoustics through a non-local active metamaterial,'' \emph{New Journal of Physics}, vol.~22, no.~6, p. 063010, 2020.

\bibitem{rosa2020dynamics}
M.~I. Rosa and M.~Ruzzene, ``Dynamics and topology of non-{H}ermitian elastic lattices with non-local feedback control interactions,'' \emph{New {J}ournal of {P}hysics}, vol.~22, no.~5, p. 053004, 2020.

\bibitem{geib2021tunable}
N.~Geib, A.~Sasmal, Z.~Wang, Y.~Zhai, B.-I. Popa, and K.~Grosh, ``Tunable nonlocal purely active nonreciprocal acoustic media,'' \emph{Physical Review B}, vol. 103, no.~16, p. 165427, 2021.

\bibitem{raval2021experimental}
S.~Raval, K.~Petrover, and A.~Baz, ``Experimental characterization of a one-dimensional nonreciprocal acoustic metamaterial with anti-parallel diodes,'' \emph{Journal of Applied Physics}, vol. 129, no.~7, p. 074502, 2021.

\bibitem{baz2022breaking}
A.~Baz, ``Breaking the reciprocity in acoustic metamaterials by active eigen-structure control strategy,'' \emph{Journal of Vibration and Acoustics}, vol. 144, no.~4, p. 041009, 2022.

\bibitem{jin2022non}
Y.~Jin, W.~Zhong, R.~Cai, X.~Zhuang, Y.~Pennec, and B.~Djafari-Rouhani, ``Non-{H}ermitian skin effect in a phononic beam based on piezoelectric feedback control,'' \emph{Applied Physics Letters}, vol. 121, no.~2, p. 022202, 2022.

\bibitem{hofmann2019chiral}
T.~Hofmann, T.~Helbig, C.~H. Lee, M.~Greiter, and R.~Thomale, ``Chiral voltage propagation and calibration in a topolectrical {C}hern circuit,'' \emph{Physical {R}eview {L}etters}, vol. 122, no.~24, p. 247702, 2019.

\bibitem{helbig2020generalized}
T.~Helbig, T.~Hofmann, S.~Imhof, M.~Abdelghany, T.~Kiessling, L.~Molenkamp, C.~Lee, A.~Szameit, M.~Greiter, and R.~Thomale, ``Generalized bulk--boundary correspondence in non-{H}ermitian topolectrical circuits,'' \emph{Nature Physics}, vol.~16, no.~7, pp. 747--750, 2020.

\bibitem{jana2023gravitational}
S.~Jana and L.~Sirota, ``Gravitational lensing and tunneling of mechanical waves in synthetic curved spacetime,'' \emph{Physical Review Research}, vol.~5, no.~3, p. 033104, 2023.

\bibitem{jana2023tunneling}
------, ``Tunneling-like wave transmission in non-{H}ermitian lattices with mirrored nonreciprocity,'' \emph{arXiv preprint arXiv:2312.16182}, 2023.

\bibitem{zhu2023higher}
P.~Zhu, X.-Q. Sun, T.~L. Hughes, and G.~Bahl, ``Higher rank chirality and non-{H}ermitian skin effect in a topolectrical circuit,'' \emph{Nature Communications}, vol.~14, no.~1, p. 720, 2023.

\bibitem{langfeldt2023controlling}
F.~Langfeldt and J.~Cheer, ``Controlling the effective surface mass density of membrane-type acoustic metamaterials using dynamic actuators,'' \emph{The Journal of the Acoustical Society of America}, vol. 153, no.~2, pp. 961--961, 2023.

\bibitem{sirota2020non}
L.~Sirota, R.~Ilan, Y.~Shokef, and Y.~Lahini, ``Non-{N}ewtonian topological mechanical metamaterials using feedback control,'' \emph{Physical {R}eview {L}etters}, vol. 125, no.~25, p. 256802, 2020.

\bibitem{zhang2021acoustic}
L.~Zhang, Y.~Yang, Y.~Ge, Y.-J. Guan, Q.~Chen, Q.~Yan, F.~Chen, R.~Xi, Y.~Li, D.~Jia \emph{et~al.}, ``Acoustic non-{H}ermitian skin effect from twisted winding topology,'' \emph{Nature {C}ommunications}, vol.~12, no.~1, pp. 1--7, 2021.

\bibitem{wen2023acoustic}
X.~Wen, H.~K. Yip, C.~Cho, J.~Li, and N.~Park, ``Acoustic amplifying diode using nonreciprocal {W}illis coupling,'' \emph{Physical Review Letters}, vol. 130, no.~17, p. 176101, 2023.

\bibitem{bender1998real}
C.~M. Bender and S.~Boettcher, ``Real spectra in non-{H}ermitian {H}amiltonians having {PT} symmetry,'' \emph{Physical {R}eview {L}etters}, vol.~80, no.~24, p. 5243, 1998.

\bibitem{supplementary}
``Supplementary material.".''

\end{thebibliography}

\onecolumngrid

\appendix

\renewcommand\x{1.6}
\renewcommand\y{3.3}
\renewcommand{\thefigure}{S\arabic{figure}}
\renewcommand{\theequation}{S\arabic{equation}}
\setcounter{equation}{0}


\newpage

\section*{\textbf{Supplementary Material}}

\section{The quantum model - Eqs. (1)-(2) of the main text}

The lattice is governed by the Hamiltonian $H_q=H^{\textrm{on}}+H^{\textrm{inter}}$, where
\begin{equation}
    H^{\textrm{on}}=i\gamma\sum_{n}(u_{nA}^{\dagger}\alpha_{nA}-u_{nB}^{\dagger}\alpha_{nB}) \quad , \quad
    H^{\textrm{inter}}=\sum_{n}(\alpha_{nA}^{\dagger}\alpha_{nB}+\eta \alpha_{nB}^{\dagger}\alpha_{(n+1)A}+h.c.)
\end{equation}
For the infinite lattice we obtain a Bloch eigenvalue problem with the effective Hamiltonian $\mathcal{H}_q$ given in Eq. (1). For $\gamma\neq 0$, $\mathcal{H}_q$ is non-Hermitian, as $\mathcal{H}_q^\dagger\neq \mathcal{H}_q$, and its eigenvalues are obtained as
\begin{equation}
\begin{split}
   & E\left[\begin{array}{c} \overline{\alpha}^A \\ \overline{\alpha}^B \end{array}\right]=\mathcal{H}_q\left[\begin{array}{c} \overline{\alpha}^A \\ \overline{\alpha}^B \end{array}\right] \quad , \quad |\lambda \textbf{I}-\mathcal{H}_q|= \left|\begin{array}{cc} \lambda+i\gamma & -f \\ -f^\dagger & \lambda-i\gamma \end{array}\right|=\left(\lambda+i\gamma\right)\left(\lambda-i\gamma\right)-ff^\dagger=\lambda^2+\gamma^2-ff^\dagger=0 \\
   & \rightarrow \quad E=\lambda=\pm\sqrt{ff^\dagger-\gamma^2}=\pm\sqrt{\eta^2-\gamma^2+2\eta\cos(ka)+1}.
\end{split}  
\end{equation}
For the PT-symmetry balance given in Eq. (2), the energies and the corresponding group velocity become
\begin{equation}
   \gamma^*=\eta-1 \quad \Rightarrow \quad E=\pm\sqrt{2\eta(1+\cos ka)} \quad , \quad v_q=\frac{\partial E}{\partial ka}=\mp \frac{\eta\sin ka}{\sqrt{2\eta(1+\cos ka)}}.
\end{equation}
To approximate $v_q$ at the vicinity of the exceptional points $ka=\pm\pi$, we expand $f$ in Taylor series up to the second order, with $\mu=ka-\pi$ defined as the nondimensional deviation wavevector. For the PT symmetric scenario we obtain the energies as
\begin{equation}
    f\approx \gamma-\tfrac{1}{2}\mu^2-i\mu \quad \rightarrow \quad E=\pm\sqrt{\gamma^2+\gamma\mu^2+\tfrac{1}{4}\mu^4+\mu^2}\approx \pm\mu\sqrt{\gamma+1} ,
\end{equation}
where it was assumed that $\mu^4/4\approx 0$. For $\gamma\gg 1$ we then obtain the energies and the group velocities in Eq. (2),
\begin{equation}
    E\approx \pm \mu\sqrt{\gamma} \quad \rightarrow \quad v_q\approx \frac{\partial E}{\partial \mu}=\sqrt{\gamma}.
\end{equation}

\section{The direct classical model - Eq. (3) of the main text}

Considering a generic mass-spring chain with masses $M_0$ and springs $K_0$, the classical equations of motion with on-site restoring forces, but complex (steady-state) gain-loss couplings, become
\begin{equation}
\begin{split}
    &\begin{cases}
    M_0\Ddot{y}^A_n&=K_0\left(y^B_{n-1}+\eta y^B_n -(1+\eta) y^A_n\right)+i\tilde{\gamma}\sqrt{K_0M_0} y^A_n \\
    M_0\Ddot{y}^B_n&=K_0\left(y^A_{n+1}+\eta y^A_n-(1+\eta) y^B_n\right)-i\tilde{\gamma}\sqrt{K_0M_0} y^B_n
    \end{cases} \\
   \rightarrow \quad  & \begin{cases}
    \omega_0^{-2}\Ddot{y}^A_n&=y^B_{n-1}+\eta y^B_n -(1+\eta) y^A_n+i\tilde{\gamma}\omega_0^{-1} y^A_n \\
  \omega_0^{-2}\Ddot{y}^B_n&=y^A_{n+1}+\eta y^A_n-(1+\eta) y^B_n-i\tilde{\gamma}\omega_0^{-1} y^B_n
    \end{cases}
\end{split}    
\end{equation}
The frequency dispersion is the same as the energy in the quantum case up to a square root of a shift, with $\Omega=\omega/\omega_0$, $\gamma=\tilde{\gamma}\omega_0^{-1}$, and $\mathcal{H}_{cd}$ defined in Eq. (3):
\begin{equation}
    \left[\mkern-8mu\begin{array}{c} y^A_n(t) \\ y^B_n(t) \end{array}\mkern-8mu\right]=\left[\mkern-8mu\begin{array}{c} \overline{y}^A \\ \overline{y}^B \end{array}\mkern-8mu\right]e^{i(kna-\omega t)} \quad \Rightarrow \quad \left(\Omega^2 \textbf{I}-\mathcal{H}_{cd}\right)\left[\mkern-8mu\begin{array}{c} \overline{y}^A \\ \overline{y}^B  \end{array}\mkern-8mu\right]=\textbf{0}.
\end{equation}

\section{The modified classical model - Eqs. (4)-(6) of the main text}

The classical equations of motion with on-site restoring forces and transient representation of the gain-loss couplings, read
\begin{equation} \label{eq:eq_motion}
\begin{split}
    &\begin{cases}
    M_0\Ddot{y}^A_n&=K_0\left(y^B_{n-1}+\eta y^B_n -(1+\eta) y^A_n\right)+\gamma\sqrt{K_0M_0} \dot{y}^A_n \\
    M_0\Ddot{y}^B_n&=K_0\left(y^A_{n+1}+\eta y^A_n-(1+\eta) y^B_n\right)-\gamma\sqrt{K_0M_0} \dot{y}^B_n
    \end{cases} \\
 \rightarrow \quad    &\begin{cases}
    \omega_0^{-2}\Ddot{y}^A_n&=y^B_{n-1}+\eta y^B_n -(1+\eta) y^A_n+\gamma\omega_0^{-1} \dot{y}^A_n \\
    \omega_0^{-2}\Ddot{y}^B_n&=y^A_{n+1}+\eta y^A_n-(1+\eta) y^B_n-\gamma\omega_0^{-1} \dot{y}^B_n
    \end{cases}
\end{split}    
\end{equation}
producing Eq. (4). 
Here, the dispersion problem is essentially different than the quantum one, taking the form
\begin{equation}
    \left[\mkern-8mu\begin{array}{c} y^A_n(t) \\ y^B_n(t) \end{array}\mkern-8mu\right]=\left[\mkern-8mu\begin{array}{c} \overline{y}^A \\ \overline{y}^B \end{array}\mkern-8mu\right]e^{i(kna-\omega t)} \quad \Rightarrow \quad \left(\Omega^2 \textbf{I}-\Omega i\gamma\bm{\sigma_z}-\mathcal{H}_0-(1+\eta)\right)\left[\mkern-8mu\begin{array}{c} \overline{y}^A \\ \overline{y}^B  \end{array}\mkern-8mu\right]=\textbf{0}.
\end{equation}
To transform from the quadratic eigenvalue problem back into a linear one, we augment the eigensystem as
\begin{equation}  \label{eq:aug}
    \left(\Omega\textbf{I}-\mathcal{H}_{cm}\right)\overline{\textbf{y}}=\textbf{0},
\end{equation}
with $\overline{\textbf{y}}$ of length 4, and $\mathcal{H}_{cm}$ defined in Eq. (5). The solution of \eqref{eq:aug} is given by
\begin{equation} \label{eq:aug_eq}
\begin{split}
   & \Omega^4-\left(2(1+\eta)-\gamma^2\right)\Omega^2+(1+\eta)^2-ff^\dagger=0 \\
   &  \Omega^2=\tfrac{1}{2}\left(2(1+\eta)-\gamma^2\right)\pm\tfrac{1}{2}\sqrt{\delta} \quad ; \quad \delta=\left(2(1+\eta)-\gamma^2\right)^2-4\left((1+\eta)^2-ff^\dagger\right).    
\end{split}    
\end{equation}
The requirements for real spectrum (with minimum at $\cos ka=-1$), therefore become
\begin{equation}
    \Omega^2\geq 0 \quad \Rightarrow \quad 1. \quad \gamma\leq \sqrt{2(\eta+1)} \quad 2.1. \quad \gamma\leq \sqrt{2}\left(\sqrt{\eta}-1\right), \quad 2.2. \quad \gamma\geq \sqrt{2}\left(\sqrt{\eta}+1\right)
\end{equation}
The condition for PT symmetry is given by 2.1, i.e. $\gamma\leq\gamma^*$, with $\gamma^*$ from Eq. (6). Then, $(1+\eta)^2-ff^\dagger=2\eta(1-\cos ka)$, $\delta=8\eta(1+\cos ka)$, and the solution of \eqref{eq:aug_eq}, as well as the corresponding normalized group velocity, read
\begin{equation}  \label{eq:v_c}
\Omega=\sqrt{\sqrt{2}\gamma+2}\cdot\sqrt{1\pm\sqrt{\tfrac{1}{2}\left(1+\cos ka\right)}} \quad , \quad v_c=\frac{\partial \Omega}{\partial ka}=\frac{1}{4}\frac{-\sin ka}{\sqrt{1+\cos ka}}\frac{\sqrt{\gamma+\sqrt{2}}}{\sqrt{\sqrt{2}\pm\sqrt{1+\cos ka}}}.
\end{equation}
At the vicinity of the exceptional points $ka=\pm\pi$, $\cos ka$ can be approximated as $-1+\mu^2/2$, with $\mu$ being the deviation wavevector, and for $\gamma\gg \sqrt{2}$, $v_c$ in \eqref{eq:v_c} thus approaches
\begin{equation}
    v_c\approx \frac{\partial \Omega}{\partial \mu}=\frac{1}{4}\sqrt{\sqrt{2}}\sqrt{\gamma}\approx 0.3\sqrt{\gamma}.
\end{equation}

\section{Numerical simulations in Fig. 2 of the main text}


\begin{figure}[htpb]
    \centering
    \setlength{\tabcolsep}{4pt}
    \begin{tabular}{c c c}
     \textbf{(a)} & \textbf{(b)} & \textbf{(c)} \\
     \includegraphics[height=4.2 cm, valign=c]{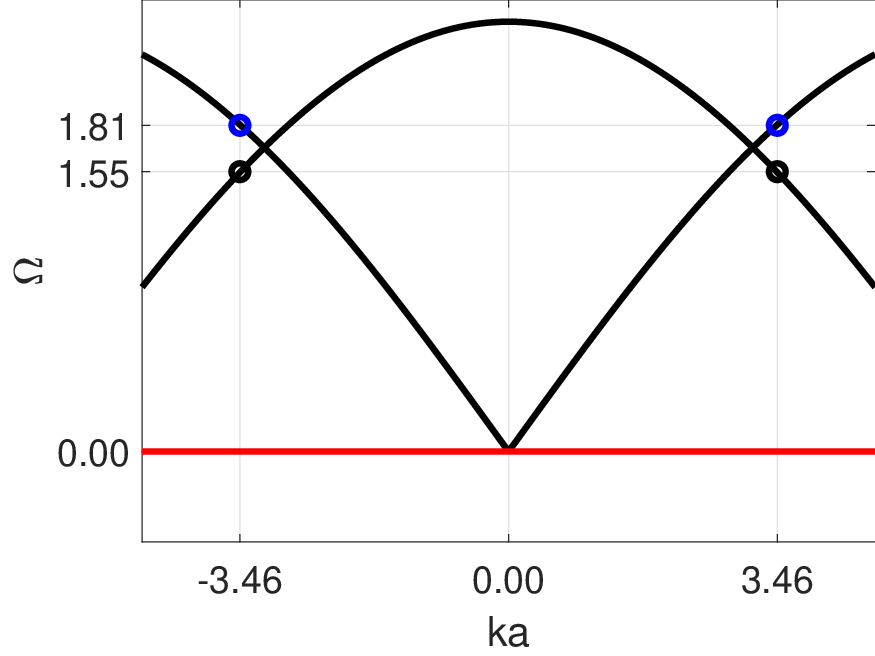}   &  \includegraphics[height=4.2 cm, valign=c]{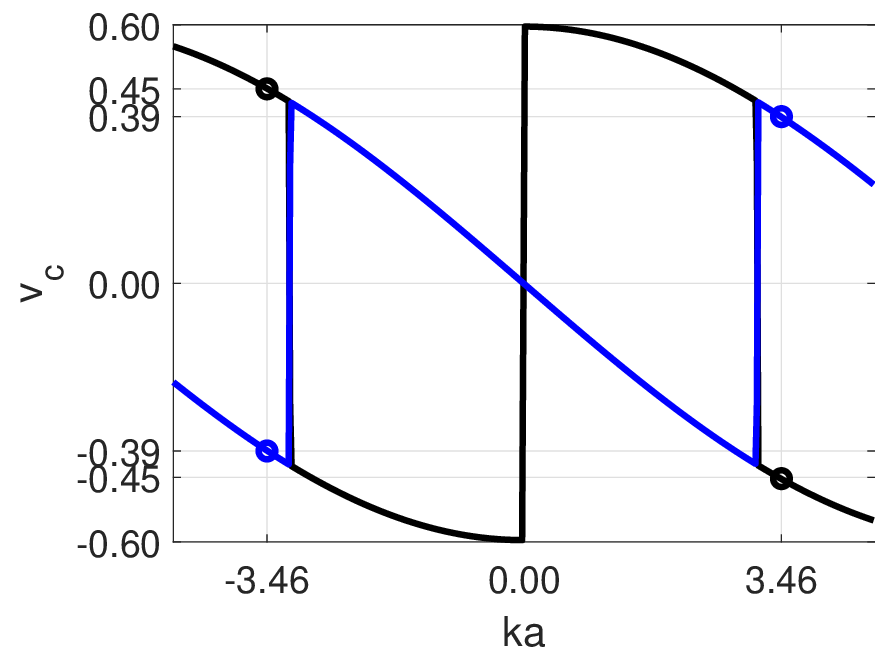}  &  \includegraphics[height=4.2 cm, valign=c]{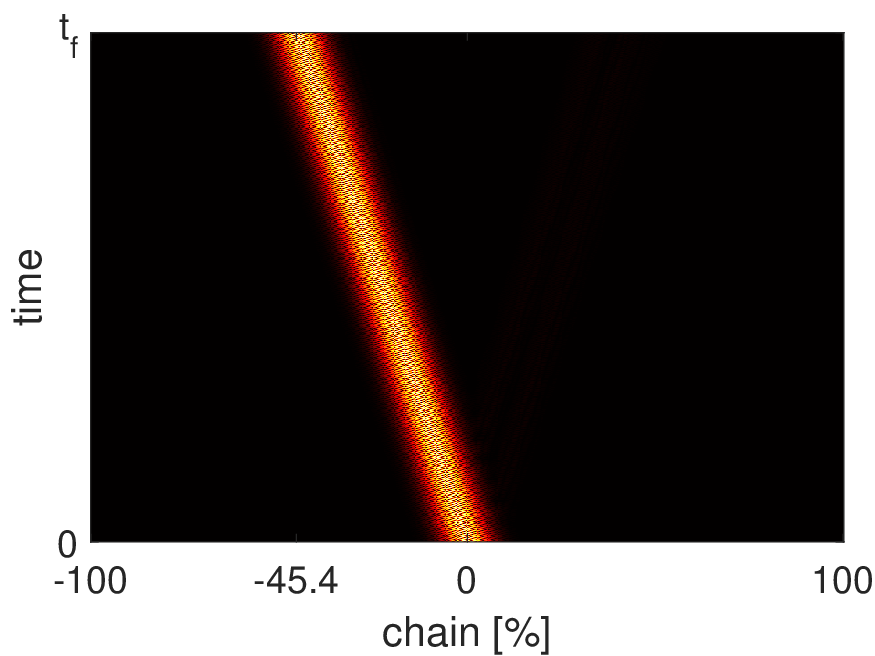}
    \end{tabular}
    \caption{\textbf{Selected wavepacket excitation for $\gamma=0.6,\eta=2$.} (a) The modified mapping dispersion, Eq. (5), marking the working points in momentum space. Each point (black, blue) corresponds to a different eigenvector. (b) The associated group velocity. (c) The time domain response at momentum $ka=3.46$ and eigenvector of $\Omega=1.55$, featuring group velocity (normalized) of -0.45 cells per second.}
    \label{fig:eigenvec}
\end{figure}

\bigskip

In order to selectively excite a particular band for a given momentum, the initial condition needs to equal the inverse spatial Fourier transform of a Gaussian in $k$ space, modulated by the corresponding eigenvector. 
Due to the bi-orthogonality property of non-Hermitian systems, the eigenvector needs to be left.
Therefore, the simulations of Fig. 2(b)-(e) were produced as follows. We defined the left eigenvector of $\mathcal{H}_{cm}$ by $\overline{\textbf{z}}$, where $\overline{\textbf{z}}^\dagger\left(\Omega \textbf{I}-\mathcal{H}_{cm}\right)=\textbf{0}$. Since $\mathcal{H}_{cm}$ is an augmented $4\times 4$ Hamiltonian, it has four complex-valued eigenvectors, each of length four. However, only two eigenvectors are relevant (the other two are idle), and for these relevant eigenvectors only two out of four entries are relevant, forming the left eigenvectors of the actual physical $A-B$ lattice. We extracted the latter using an appropriate sorting, and denoted them by $\overline{\textbf{z}}_1^\dagger=[\begin{array}{cc}
   \overline{\textbf{z}}^A_1  &  \overline{\textbf{z}}^B_1
\end{array}]$ and $\overline{\textbf{z}}_2^\dagger=[\begin{array}{cc}
   \overline{\textbf{z}}^A_2  &  \overline{\textbf{z}}^B_2
\end{array}]$, where
\begin{equation}
    \overline{\textbf{z}}^A_1=r_1^Ae^{i\phi_1^A} \quad , \quad \overline{\textbf{z}}^A_2=r_2^Ae^{i\phi_2^A} \quad , \quad \overline{\textbf{z}}^B_1=r_1^Be^{i\phi_1^B} \quad , \quad \overline{\textbf{z}}^B_2=r_2^Be^{i\phi_2^B}.
\end{equation}
Then, in order to excite a selected momentum $ka$, we defined the initial condition for the $y$ and $\dot{y}$ degrees of freedom, as
\begin{equation}  \label{eq:IC}
    \left[\mkern-8mu\begin{array}{c} y^A_n(0) \\ y^B_n(0) \end{array}\mkern-8mu\right]=G_n\left[\mkern-8mu\begin{array}{c} r_j^Ae^{i\phi_j^A} \\ r_j^Be^{i\phi_j^B} \end{array}\mkern-8mu\right]e^{inka} \quad , \quad \left[\mkern-8mu\begin{array}{c} \dot{y}^A_n(0) \\ \dot{y}^B_n(0) \end{array}\mkern-8mu\right]=-i\omega G_n\left[\mkern-8mu\begin{array}{c} r_j^Ae^{i\phi_j^A} \\ r_j^Be^{i\phi_j^B} \end{array}\mkern-8mu\right]e^{inka} \quad , \quad G_n=e^{-\frac{(na-\delta)^2}{\sigma^2}} ,
\end{equation}
where $G_n$ is a Gaussian envelope for real space localization. The index $j=1,2$ stands for each of the two eigenvectors, respectively corresponding to the top (blue circle) and bottom (black circle) dispersion band, Fig. \ref{fig:eigenvec}(a). 
The actual initial conditions must be real, and thus a possible way of representing \eqref{eq:IC}, is
\begin{equation}  \label{eq:IC_real}
    \left[\mkern-8mu\begin{array}{c} y^A_n(0) \\ y^B_n(0) \end{array}\mkern-8mu\right]=\left[\mkern-8mu\begin{array}{c} r_j^A\sin (kna+\phi_j^A) \\ r_j^B\sin (kna+\phi_j^B) \end{array}\mkern-8mu\right] \quad , \quad \left[\mkern-8mu\begin{array}{c} \dot{y}^A_n(0) \\ \dot{y}^B_n(0) \end{array}\mkern-8mu\right]=-\omega\left[\mkern-8mu\begin{array}{c} r_j^A\cos (kna+\phi_j^A) \\ r_j^B\cos (kna+\phi_j^B) \end{array}\mkern-8mu\right].
\end{equation}
The simultaneous initial condition of $y$ and $\dot{y}$ respectively correspond to the sine and cosine components of the inverse Fourier transform, which means capturing only the actual momentum $ka$ without its counterpart $-ka$. The consequence for time domain is a unidirectional propagation of the wavepacket. 
For example, exciting the bottom band of momentum $ka=3.46$ (the black point in Fig. \ref{fig:eigenvec}(a)) produces a wavepacket that propagates with a normalized velocity of -0.45, as follows from the group velocity plot in Fig. \ref{fig:eigenvec}(b). Indeed, in the time domain simulation in Fig. \ref{fig:eigenvec}(c), the wavepacket travels to the left with the required velocity. Its counterpart that corresponds to the top band (the blue point), traveling to the right with the velocity 0.39, appears in Fig. 2(c) of the main text. The other responses in Fig. 2 were produced in a similar way.

\section{Topoelectric metamaterial realization in Fig. 3 of the main text}

\begin{figure}[htpb]
    \centering
    \includegraphics[height=4.5 cm]{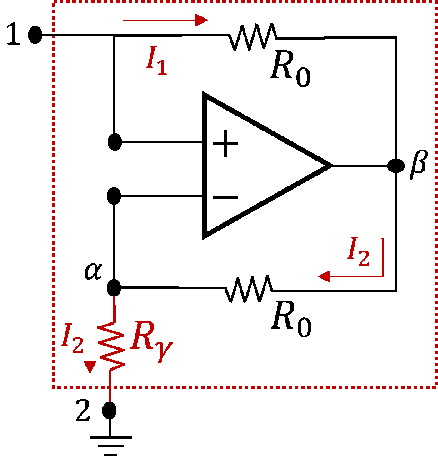}
    \caption{\textbf{Realization of the gain $\gamma$ via negative resistance $-R_\gamma$ using an operational amplifier.}}
    \label{fig:gain}
\end{figure}

Defining $M_0=C_0$ and $K_0=1/L_0$ for a capacitor $C_0$ and inductor $L_0$, we obtain $\omega_0^2=1/(L_0C_0)$. The governing equations are then given by \eqref{eq:eq_motion} with the voltage $V$ instead of displacement $y$. For the first chain site, the equation of motion includes also the source term $V_s$ and the signal generator resistance $R_g$,
\begin{equation}
    \omega_0^{-2}\Ddot{V}^A_n+\frac{L_0}{R_g}\dot{V}^A_n=V^B_{n-1}+\eta V^B_n -(1+\eta) V^A_n+\gamma\omega_0^{-1} \dot{V}^A_n+\frac{L_0}{R_g}\dot{V}_s(t).
\end{equation}
The gain of the $A$ sites is realized by an operational amplifier in a negative impedance converting setup, as follows from Fig. \ref{fig:gain}, which is the detailed version of Fig. 3(b). 
We have $I_1=(V_1-V_\beta)/R_0$ and $I_2=(V_\beta-V_\alpha)/R_0$, which due to $V_1=V_\alpha$ gives the expected $I_2=-I_1$. 
Since we also have $I_2=(V_\alpha-V_2)/R_\gamma$, where $R_\gamma=\sqrt{C_0/L_0}\gamma$, we obtain
\begin{equation} \label{eq:V_B}
    V_\beta=V_1\left(1+\frac{R_0}{R_\gamma}\right)-V_2\frac{R_0}{R_\gamma}.
\end{equation}
Equating \eqref{eq:V_B} with $V_\beta=-R_0I_1+V_1$, leads to
\begin{equation} \label{eq:V_12}
    V^A_n=V_1-V_2=-R_\gamma I_1.
\end{equation}
On the other hand, the loss of the $B$ sites is realized by a regular resistor, so that
\begin{equation}
    V_n^B=+R_\gamma I.
\end{equation}

\end{document}